# Nanoscale Structure and Elasticity of Pillared DNA Nanotubes

**Himanshu Joshi[1], Atul Kaushik[2], Nadrian C. Seeman[3] and Prabal K. Maiti[1*]**

[1]Center for Condensed Matter Theory, Department of Physics, Indian Institute of Science, Bangalore 560012, India

[2]Department of Biotechnology, Indian Institute of Technology Madras, Chennai 600 036, India

[3]Department of Chemistry, New York University, New York, NY 10003, USA

*Address the correspondence to maiti@physics.iisc.ernet.in

**ABSTRACT**

We present an atomistic model of pillared DNA nanotubes (DNTs) and their elastic properties which will facilitate further studies of these nanotubes in several important nanotechnological and biological applications. In particular, we introduce a computational design to create an atomistic model of a 6-helix DNT (6HB) along with its two variants, 6HB flanked symmetrically by two double helical DNA pillars (6HB+2) and 6HB flanked symmetrically by three double helical DNA pillars (6HB+3). Analysis of 200 ns all-atom simulation trajectories in the presence of explicit water and ions shows that these structures are stable and well behaved in all three geometries. Hydrogen bonding is well maintained for all variants of 6HB DNTs. We calculate the persistence length of these nanotubes from their equilibrium bend angle distributions. The values of persistence length are ~10 μm, which is 2 orders of magnitude larger than that of dsDNA. We also find a gradual increase of persistence length with an increasing number of pillars, in quantitative agreement with previous experimental findings. To have a quantitative understanding of the stretch modulus of these tubes we carried out nonequilibrium Steered Molecular Dynamics (SMD). The linear part of the force extension plot gives stretch modulus in the range of 6500 pN for 6HB without pillars which increases to 11,000 pN for tubes with three pillars. The values of the stretch modulus calculated from contour length distributions obtained from equilibrium MD simulations are similar to those obtained from nonequilibrium SMD simulations. The addition of pillars makes these DNTs very rigid.

DNA is an excellent material of choice for the precise organization of chemical species on the nanoscale.[1-4] The specific base pairing rules that enable DNA to function as a biological information carrier also facilitate the construction of DNA assemblies in a programmable manner. The interstrand interactions and sticky ended cohesion of DNA molecules provide a scaffold for the synthesis of branched DNA nanostructures with excellent precision.[5] In the past few decades, various novel nanostructures such as cubes,[6] tetrahedra,[7] truncated octahedra,[8] icosahedra,[9] smiley faces, two dimensional arrays[10-13] and nanotubes[14-17] have been synthesized using the versatile DNA molecule. Nanomechanical devices have also been constructed using DNA motifs, extending their potential application to DNA nanorobotics.[18-20] The introduction of DNA origami[21] has significantly accelerated the synthesis of various DNA nanostructures. Looking forward to the application of DNA nanotubes (DNTs) in cellular drug delivery, researchers are investigating their self-assembly and interaction with lipid membranes.[22, 23] The development of techniques such as DNA origami along with computational tools such as caDNAno[24] and 3DNA[25] has made their fabrication and analysis fast and efficient. With an ever-increasing repertoire of possible structures that can be constructed using DNA, it becomes imperative to characterize their behavior in solution.[26, 27] This would help us to understand the effectiveness of the fabrication process and offer insights into the performance and construction of complex assemblies in a modular fashion. Experimental techniques that have been used to study DNA assemblies include atomic force microscopy,[14, 28] magnetic tweezers,[27] cryo-electron microscopy[29] and X-ray crystallography[30] to name a few. For a complete understanding of their solution behavior, static techniques such as those mentioned above must be supplemented by techniques such as fluorescence spectroscopy[31] and molecular dynamics[32, 33] that probe the dynamics of these assemblies.

DNTs are self-assembled molecular nanopores with a programmable circumference.[34] These nanotubes are made of DNA tiles which can readily self-assemble in one dimensional arrays up to 10 μm, with a variable diameter exploiting the hybridization of sticky ends.[35] Here, each unit of DNT has at least *2n* immobile crossovers or Holliday like junctions[36] between the strands of neighboring double-stranded (dS) DNA, where *n* is the number of helical domains. It has been shown that juxtaposed dsDNA can exchange strands at various points *via* reciprocal exchange, producing stiff motifs.[35, 37,38] These DNTs can also be viewed as the extension of the DX,[39] TX,[40] or PX[41] crossover molecule motifs in a closed circular geometry. Single Holliday junctions are very flexible and it is very important to design the crossover points between the helices in order to make the nanotube robust and stable.[42] Studies have revealed that with chemical ligation, it is possible to enhance the stability and mechanical strength of DNTs.[43] DNTs with enhanced rigidity, have proposed applications in nanoelectronics,[44] nanomedicine,[45] nanomechanical[20] devices and various biophysical studies.[46] With better design and assembly techniques, they may be able to mimic synthetic microtubules.[47] Recently, Wang *et al.* experimentally characterized DNTs and studied the change in mechanical properties of a six helix nanotube upon the addition of flanking helices that we term 'pillars'.[14] The authors synthesized one dimensional DNTs (with and without pillars) by the addition of sticky ends at the ends of the monomers. The end-to-end distances and contour lengths of these DNTs were extracted from fluorescence images and fitted to the Worm-like chain (WLC) model to obtain their persistence lengths. It was observed that there was an increase in rigidity with the addition of pillars.[14] In this work, we aim to characterize the structure, dynamics and rigidity of these nanotubes (with and without pillars) that have parallel ends using state-of-the art Molecular Dynamics (MD) simulations. In recent years although several reports have appeared explaining the experimental synthesis and characterization of various DNTs,[15-17, 42, 44, 48] the molecular level characterizations of pillared DNTs are still lacking. Recently, a few groups have approached the study of self-assembled DNA nanostructures using atomistic as well as coarse grained simulations.[33, 49-51] Maingi *et al.* have shown the gating like behavior of DNTs using molecular simulations.[52] Atomistic molecular dynamics simulations are likely to provide useful insights for the transition of DNTs from the laboratory to their practical and industrial implementation in day to day life.

In this work, we study the microscopic details of the structural and dynamical aspects of 6HB, 6HB+2 and 6HB+3 DNTs using all atom MD simulations. We have come up with an in-house code to build an atomistic model (Figure 1) of these nanotubes. The rigidity of DNTs is an important property which influences their interaction with their environment. Early experiments with DNTs report persistence lengths of several micrometer.[17, 43] We attempt to estimate and compare the persistence lengths of DNTs using MD simulation techniques. The stiffness of the components used in DNA nanotechnology is of key interest. The essence of DNA is that it provides a nanoscale avenue for building objects, lattices and nanomechanical devices by programming the sequences of its constructs.[2] One would like to be able to treat these species like classical building materials, so that analogs of bricks, beams, hinges and the like behave as they do on the macroscopic scale. In the absence of this behavior, the DNA nanotechnology would be much less reliable. The rest of the paper is organized as follows: First, we discuss the structural and dynamical analysis of the simulation data including the Root-Mean-Square-Deviation (RMSD), Root-Mean-Square-Fluctuation (RMSF) in atomic positions, radius analysis of the pore and interstrand hydrogen bonding profiles of these DNTs. Next, we present the calculation of the stretch modulus and the persistence length using equilibrium and non-equilibrium MD simulations. Then, we conclude the current study with emerging future directions. Finally, details of the DNT building protocol and the simulation methodology have been provided towards the end of the paper.

**RESULTS AND DISCUSSION**

**Structural Deformation: Time Evolution of RMSD and RMSF.**

Figure 2 (a) shows the RMSD of various DNTs with respect to their minimized structure. While calculating the RMSD, only the atoms belonging to the core 6 helix bundle (without the flanking dsDNA) were taken into account. The RMSD values were averaged over three sets of simulations. Over several nanoseconds time scale, the RMSD values settled at ~ 15 Å for all the three structures indicating convergence to equilibrium solution structures. The RMSD value for the 6HB+2 is marginally lower compared to 6HB+3 followed by 6HB. The RMSD per dsDNA helix is in the range of 2.5 Å which is normally observed for solution state simulations of regular dsDNA. Instantaneous snapshots of all three structures after 200 ns shown in Figure 3 confirm that tubular structures are maintained for all of the cases and show the molecular level details of various nanotube topologies. The average geometrical DNT parameters after 200 ns of the base pairs, base steps and helices are given in Tables 2a-c, respectively. The parameters were calculated using CPPTRAJ by following the 3DNA protocol.[25] Comparison with the initially built structure shows that the after 200 ns MD simulation, the structures preserve their basic DNA parameters. Video V2 (in Supporting Information SI) shows a movie of the trajectory of the 200 ns MD simulation of the 6HB DNT. The evolution of RMSD with respect to the initial minimized structure is shown in video V3. To account for the fluctuations in atomic positions, we have plotted the average RMSF for different cross sections along the length of the axis of the tube. 57 bp long DNTs were divided into the 57 slices, taking one bp from each helical domain, as shown in Figure S4 in the SI. The structure was fitted to the initially energy minimized structure and then RMSF of the atoms in particular slice has been calculated for the 200 ns trajectory. Figure 2 (b) shows the RMSF values of the corresponding slice averaged over three sets of simulations for each construct. The center region of the nanotubes is more stable than the outer edges in all the three cases. The values of RMSF are around 14 Å at both the ends of nanotube and symmetrically decrease to 4 Å at the center region of the DNT. Overall, the RMSF for the central core of all three kinds of nanotubes is similar within the error bars. This analysis shows that the core of all the three types of DNTs is equally stable. The RMSD and RMSF analyses for full 6HB+2 and 6HB+3 DNTs are given in Figure S2 in SI.

**Radius of the Pore along the Helical Axis.**

The pore of the DNT is a key feature for the application of these structures in nanotechnology. We calculate the radius of this hollow channel of DNT using the simulated structures. The radius profile is computed using the same technique used in our previous work.[32] To calculate the radius profile, we divide the nanotube into 1 Å small segments across the helical axis of the DNT (as described in figure S3 of the SI). The radius of the section is computed by calculating the root mean squared distance from that center as described by the following equation

$$R = \frac{1}{n}\sum_{i=1}^{n}\sqrt{\left(x_i - x_c\right)^2 + \left(y_i - y_c\right)^2} \qquad (1)$$

The radius profiles of the three different types of DNTs are shown in Figure 4. The data were averaged over the last 10 ns of MD simulations from the three independent 200 ns MD simulation for each DNT. The plot shows that both the ends of DNTs are wide open with a radius of ~3 to 3.5 nm while the center region of the pore maintains a circular geometry with a radius of 2.5 nm. This is in quantitative agreement with the experimental and theoretical expectations.[14, 16] The widening of the nanotubes at both the ends can also be seen from the instantaneous snapshots of DNTs as shown in Figure 3. The geometrical parameters show more fluctuations at the termini. The radius fluctuations at both ends of DNTs are due to the fraying of base pairs and the finite size effects of DNTs. In experiments where we have many such units of DNTs, we expect a smoother radius profile with a radius of 2.5 nm. The radius profile including the outer helices is given in Figure S6 in SI.

**Hydrogen Bond Analysis and Broken Base Pairs.**

Hydrogen bonding interactions between the base pairs play a vital role in the stabilization of nucleic acids in solution. Consequently, we analyzed the dynamics of the hydrogen bonds between the base pairs of DNTs to assess the stability of these structures. The variation of the broken hydrogen bonds between the base pairs as a function of the simulation is shown in figure 5. We used criteria similar to those suggested in the IUPAC reference to define hydrogen bonds.[53] In particular, we have used a distance cutoff of 3.5 Å and an angle cut-off of 120°. The percentage of broken hydrogen bonds varies from zero, at an initial stage of simulation, to ~ 6% after a 200 ns production run. The broken base pairs were also averaged over all three sets of simulations. The fraction of broken base pairs saturates after the initial equilibration to an almost constant value for all three structures. On the basis of the fraction of broken hydrogen bonds between the base pairs, the core of all three structures looks equally stable. We observe that the major contribution to broken hydrogen bonds comes from the terminal base pairs of the DNTs. To quantify the contribution from the various regions of DNTs, we plot the dynamic correlation between the segment-wise average number of broken base pairs and the simulation time in figure

6. Here, we divide the DNTs into nine symmetrical segments as described in figure S4 (SI) and plot the fraction of the average broken hydrogen bonded base pairs for that segment in color code. The terminal segments (segments 9 and 1) have the most broken base pairs. The terminal base pairs of DNA are expected to undergo frequent opening ("fraying").[54] The plot also suggests that the regions away from the crossovers show more fluctuation. Note that, only the core 6HB helices of 6HB+2 and 6HB+3 were considered in these calculations. The broken base pairs analysis for whole 6HB+2 and 6HB+3 nanotubes are shown in Figure S5 in SI. Figure S5b in SI compares the fraction of broken hydrogen bonds in the pillars of 6HB+2 and 6HB+3 structures.

**Mechanical Properties.**

With the advancement of single molecule manipulation techniques, the elastic properties of DNA molecules were studied extensively using several experimental techniques like magnetic tweezers, optical tweezers and atomic force microscopy.[55-60] The force *versus* extension curves from these experiments are normally fitted by different versions of the freely jointed chain model (FJC)[61] or the WLC[62] to extract the persistence length and the stretch modulus of DNA structures. After the discovery of DNTs, there was continuous interest to measure and find ways to enhance the mechanical strength of these nanotubes, as reviewed recently by Castro *et al*.[63] The pillared nanotubes 6H+2 and 6H+3 were proposed to have greater stiffness in terms of persistence length compared to the simple 6HB molecule.[14] Earlier we have reported the persistence length and stretch modulus of various 6HB nanotubes without pillars, using force-extension behavior from steered molecular dynamics (SMD)[32] simulations Here, we extend our effort to measure the persistence length of the pillared nanotubes using two different approaches: from the force extension behavior and from the analysis of the equilibrium bend angle distribution and contour length distribution. We have recently studied the mechanical strengths of DNA in various conformations of dsDNA using similar techniques.[46, 64]

**Force-Extension Behavior.**

We have performed nonequilibrium SMD[65] simulations under constant velocity ensembles for various DNTs. The velocity applied for the pulling simulation was 1Å/ns. In order to sample phase space better, we performed two sets of pulling simulations using different initial conditions. Video V4 in SI shows a movie for one such pulling simulation of the 6HB DNT (for clarity water and ions are not shown in the movie). Figure 7 shows the average force-extension behavior from both sets of pulling runs. The force-extension curve starts with the initial entropic region where the extension is negligible compared to the contour length of DNTs. This region is followed by the linear elastic region where the DNTs have been extended by up to 10% of their equilibrium contour length. In this region, the hydrogen bonds between the base pairs are intact and the backbone is slightly deformed. Beyond this region, we see a plateau region in the curve where the DNTs are stretched up to ~30% of their initial contour length. This overstretched region with a small force variation corresponds to the B-to-S transition in dsDNA.[66, 67] We have extracted the stretch modulus of DNTs from the linear fit of the elastic region using the Hooke's law. The values of stretch moduli of 6HB, 6HB+2 and 6HB+3 are **6533.3**(± 355.4) pN, **8553.4** (±364.6) pN and **10932.3** (±317.3) pN, respectively. With increasing numbers of pillars, the stretch modulus increases and the 6HB+3 DNT has the highest stretch modulus.

**Contour Length Distribution and Stretch Modulus.**

Small fluctuations in the equilibrium contour length of DNTs can account for their stretch moduli. In this model, we assume the DNTs behave as elastic rod. In this case, a small fluctuation, $\Delta L$, around the equilibrium contour length, $L_0$, of the DNT will generate a restoring force $F = -\gamma \, \Delta L/L_0$, where $\gamma$ is the stretch modulus of the DNTs. In a canonical ensemble at constant temperature $T$, the probability of having an instantaneous contour length L can be given as

$$P(L) = \sqrt{\frac{\gamma L_0}{2 \pi k_B T}} e^{-\frac{\gamma L_0}{2 k_B T}(L/L_0 - 1)^2} \qquad (2)$$

$$\Rightarrow \quad \ln P(L) = -\frac{\gamma L_0}{2 k_B T}(L/L_0 - 1)^2 + C. \qquad (3)$$

Here, the contour length of DNTs is taken to be equal to the average of the contour length of the individual helices. We took the snapshots of the last 100 ns (i.e., 100−200 ns) from three independent simulations to compute the distribution of contour lengths for a given system. The normalized distributions along with fits to the Gaussian formula as given by Eqn. 2 are shown in Figure 8a. We extract the stretch modulus by fitting a linear regression using equation 3 as shown in figure 8b. The estimated stretch moduli of 6HB, 6HB+2, 6HB+3 are **6030.8** (±309.8) pN, **9765.2** (±291.8) pN and **13938.6** (±461.2) pN respectively. The values of the stretch modulus from this analysis are in good agreement with the values obtained from force extension behavior.

**Persistence Length from Bending Angle Distribution.**

We have also calculated the persistence length of these nanotubes from the bend angle distribution derived from the equilibrium simulation.[67-70] Recently, we investigated the flexibility of ds-DNA at different salt concentrations using a similar analysis.[71] Here we briefly give the details of the analysis. The bending of DNA can be characterized by the bending angle $\theta$. The probability of finding a small fluctuation $\theta$ in the bending angles of a flexible polymer in a canonical ensemble at temperature $T$, can be written as

$$P(\theta) = \sqrt{\frac{\kappa}{2\pi k_B T L_0}} e^{\frac{-\kappa}{2 L_0 k_B T}\theta^2} \quad (4)$$

$$\Rightarrow \ln P(\theta) = -\frac{L_p}{L_0}(1 - \cos\theta) + C \quad (5)$$

Where $L_0$ is the average contour length, $\kappa$ is the bending modulus, $k_B$ is the Boltzmann constant and $L_p = \kappa/k_B T$ is the persistence length.

To define the bending angle, the DNT was divided into nine sections, where the terminal sections, one and nine, consist of 4 base pairs and the remaining inner sections contain seven base pairs each (as described in figure S5 in the SI). The center of mass (COM) of each of the sections was calculated from the simulation trajectories. The tangent at section $i$ was defined as the line connecting the COM of section $i$ with that of section $i+1$. Tangents for sections one to eight were defined in this manner. Both the ends of DNTs fluctuate extensively, so we neglected both the end segments to calculate the bend angle distribution. The last 100 ns of the trajectories from the three independent simulations were used to calculate the bend angle distribution of each DNT, shown in Figure 9a. From the fit of the bend angle distribution as shown in figure 9b, we calculate the persistence length as given by equation 5. The persistence lengths measured for 6HB, 6HB+2 and 6HB+3 are **1.8** (±.1) μm, **2.6** (±0.1) μm and **2.9** (±0.1) μm respectively. Table 3 summarizes the results extracted from the above analysis. Our simulation results are in quantitative agreement with the experimentally measured persistence lengths.[14]

**Persistence Length from Stretch Modulus.**

The persistence length of the DNT can also be calculated using elastic theory, assuming the nanotube to be an elastic rod, using the following expression

$$L_p = \frac{EI}{k_B T} = \frac{\kappa}{k_B T} \qquad (6)$$

Where $E$ is the Young's modulus of the DNT which is given by $E = \frac{S}{Area}$, $S$ is the stretch modulus, $I$ is the area moment of inertia (AMI) of the structure, $k_B$ is the Boltzmann constant and $T$ is absolute temperature.

Substituting the calculated values of the stretch modulus from force-extension behavior and contour length fluctuations in equation 6, we estimate the persistence length $L_p$ of the DNTs. AMI has been calculated by averaging two bending axes of the DNTs as shown in figure S7 (for details refer to the SI). We approximate the individual DNA helix as an isotropic elastic rod of radius 1 nm. Following our radius analysis, the radii of the inner and outer cores of the DNT has been taken to be 2.5 nm and 5.0 nm, respectively.

The persistence length of 6HB, 6HB+2 and 6HB+3 are **5.3(±0.3) μm, 12.1(±0.5) μm** and **15.2(±0.5) μm** respectively when we use the values of the stretch modulus obtained from the force extension analysis. Using the stretch modulus values obtained from the contour length distribution, we get persistence lengths **4.9**(±0.2) μm, **13.5**(± 0.4) μm and **21.9**(±0.7) μm for 6HB, 6HB+2 and 6HB+3 respectively. Table 4 summarizes the values of stretch moduli and corresponding persistence lengths using the stretch modulus analysis. Although, the numbers are much higher than those obtained from the bend angle distribution analysis, the trend is similar. Keeping in mind that we approximated the individual DNA helix as an isotropic rigid rod for the calculation of AMI, higher persistence lengths from the stretch modulus are expected. We see the swelling of DNTs in the course of simulation which gives rise to a higher radius compared to an ideal DNT. This leads to higher values of the AMIs and hence the larger persistence lengths. Figure 10 compares the persistence lengths and stretch modulus of DNTs obtained from different simulation approaches and the experimental persistence length reported by Wang *et al.*[14] As the structures have some nicks in the backbones, the respective base-pairs will be prone to deviate from ideal base stacking and base pairing giving rise to a decrease in mechanical strength. This is also evident from the movie of the pulling simulation (video V4 in the SI). Thus, using a variety of simulation techniques and analyses, we have demonstrated an increase in the rigidity of DNTs with the addition of pillars. Owing to assumptions in the calculations involved, we believe that the trends obtained carry greater significance as compared to the exact numerical values of the mechanical parameters. The results also validate the analyses protocols used in a self-consistent fashion.

**CONCLUSIONS**

The structural and dynamical study of 6HB, 6HB+2 and 6HB+3 DNTs suggests that, apart from the fraying of base pairs and back-bone deformation at both termini, these structures are stable in physiological conditions. The RMSD and RMSF calculations quantify the conformational fluctuations during the 200 ns long MD simulation. The fluctuations are more prominent at the terminals of nanotubes which could be a consequence of the finite size of these simulated DNTs. The critical observation regarding base pair dynamics reveals that ~ 95% of the hydrogen bonds are intact during the course of the MD simulation. Visual examination of the simulation snapshots and the analysis of geometrical parameters shows that base stacking is well maintained in spite of the backbone geometry being slightly deformed. Based on these analyses, we believe that stacking interactions and intact hydrogen bonds between the base pairs provide crucial stability to these DNA crossover nanostructures. The radius analysis shows that despite the large fluctuations on both ends, the DNTs maintain a tubular cross-section in their central regions, with a radius of 2.5 nm, including the van der Waals radii of the DNA atoms. We have characterized the mechanical strengths of these DNTs using their force-extension behavior in addition to analysis of their equilibrium conformational analysis. With an increasing number of pillars, the DNTs become stiffer and have progressively larger stretch moduli and longer persistence lengths.

Adding pillars emerges as a natural and experimentally convenient way to increase the stiffness of nanoscale pores. Overall, we find our simulation results match qualitatively with recent experimental findings.[14, 17] The DNTs have persistence lengths of ~ 10 μm which approaches the ballpark of the persistence length of microtubules *i.e.* ~ 1000 μm. This provides optimism that with better experimental design and understanding of the self-assembly, synthetic crossover DNA nanostructures can potentially target the mechanical behavior of microtubules. The evident suggestion is that larger numbers of parallel helices flanking a central cavity, already seen in various origami constructs, ought to provide the basis for very stiff nanotubes. This study has helped to understand the microscopic picture of DNTs and has given useful insights for the improved design of DNA crossover nanostructures. With the recent progress in the coarse grained representation of DNA,[72] larger DNA nanostructures can be studied at longer time and larger length scales, thereby providing more elaborate insight about the bottom up synthesis of DNTs.

## METHODS

**Design and Construction.**

The initial structures for the DNTs were designed using a custom built program written in NAB,[73] an AMBER[74] utility to construct DNA nanostructures. The structure was decomposed into helical segments that were placed at their appropriate positions to form the intended shape/topology. These segments were then connected *via* phospho-diester bonds to get the final structure. The sequence used for the inner core of the structures is given in figure S1 in the SI. The pillars were attached to the inner core by crossovers spaced by 21 base pairs (bp). The base pair spacing is governed by the hexagonal geometry of inner core. This facilitates crossovers among the helical domains after a separation of 7 bp (240°) or 14 bp (120°) or those angles plus an exact multiple of 360°. The structure also has nicks in the backbone of the double helical domain represented by in figure S1. The crossovers and nicks in the structures are identical to the experimental designs by Seeman *et al.*[14, 15] Figure 1 shows the images of the NAB built structures of DNTs. These images have been generated using VMD.[75] A rotating 3D view of initial build 6HB DNT using PyMol is shown in the video V1.[76]

**Simulation Methodology.**

The output pdb structure of the DNTs from the custom made NAB code was loaded in the xleap module of AMBER.[74, 77] AMBERff99 force fields[78, 79] with parambsc0 corrections[80] have been used to describe the bonded and non-bonded interactions involving the DNA atoms. The structure was immersed in the rectangular box of the TIP3P[81] water model. We ensured a solvation shell of 15 Å around the DNT. Subsequently, the appropriate numbers of Na+ ions were added to neutralize the system's charge balance. The Joung-Chetham (JC) ion parameters were used to model Na+ ion's interactions with DNA and water.[82] The ions were placed around the DNT by constructing a Coulombic potential grid of 1 Å and then placing the ion at a site of highest electrostatic potential. The details of all the simulated systems are given in Table 1. The systems were then subjected to energy minimization to eliminate bad contacts of the DNT with water and ions. During the course of energy minimization, we restrained the solute using harmonic constraints, which were gradually reduced from 500kcal/mol-Å$^{-2}$ to 0 in several thousand steepest-descent and conjugate-gradient minimization steps. The system was then slowly heated up to 300 K in 40 ps with 1 fs integration time step. The solute was kept fixed during heating with a weak harmonic constraint of 20 kcal/mol-Å$^{-2}$. We then equilibrated the structures at 1 atm pressure and 300 K temperature to get the correct density for 100 ps. Finally, 200 ns long MD production runs were carried out in a canonical ensemble with an integration time step of 2 fs. Particle mesh Ewald (PME) technique integrated with amber molecular dynamics package was used to treat the long range interaction with a 10 Å cut off for the short range non-bonded interactions.[83, 84] The covalent bonds involving hydrogen were constrained using the SHAKE[85] algorithm to increase the integration time step. A Berendsen weak coupling thermostat was applied to maintain the temperature constant at 300° K with a coupling constant of 0.5 ps. The translational motion of the center of mass (COM) has been removed after 1000 time steps. Similar simulation protocol was successfully validated and used in our previous MD simulation studies on DNA nanostructures.[86-89] We have carried out three sets of statistically independent simulations for each type of DNTs using the above simulation protocol.

To calculate the elastic properties of these DNTs, we have performed steered molecular dynamics simulations (SMD) in constant velocity ensembles. For the pulling simulation, we have applied a force on the O3' and O5' atoms of all the single strands on both ends of the DNT. The force has been applied along the vector joining the O3' and O5' atoms of the terminal strands. We pull the DNTs using constant velocity simulation with a velocity of 1 Å/ns which is several orders of magnitude faster than the typical experimental pulling velocities used in experiments. Studies have shown that the faster pulling rate requires a higher force to see the same strain in the structures compared to a lower pulling rate.[90] We tested our simulation with the slower pulling rate (0.5 Å/ ns) which gave similar force-strain characteristics of DNTs but with lower force values for various regions in the force-extension curve. We have extensively studied the mechanical properties of B-DNA and its variant using a similar pulling protocol.[32, 87, 91] CPPTRAJ,[92] an AMBERTOOLS[74] utility has been used extensively in various calculations presented above.

**Acknowledgement.** We thank Department of atomic energy (DAE) and DST, India for financial assistance, Supercomputer Education and Research Centre (SERC), IISc for providing supercomputer facilities specially SahasraT. HJ thanks to CSIR for the senior research fellowship. NCS has been supported by grants EFRI-1332411, and CCF-1526650 from the NSF, MURI W911NF-11-1-0024 from ARO, MURI N000140911118 from ONR, DE-SC0007991 from DOE for partial salary support, and grant GBMF3849 from the Gordon and Betty Moore Foundation.

# REFERENCES.

1. Seeman, N. C. Nucleic Acid Junctions and Lattices. *J. Theor. Biol.* **1982**, 99, 237-247.

2. Seeman, N. C. DNA in a Material World. *Nature* **2003**, 421, 427-431.

3. Modi, S.; Bhatia, D.; Simmel, F. C.; Krishnan, Y. Structural DNA Nanotechnology: From Bases to Bricks, from Structure to Function. *J. Phys. Chem. Lett* **2010**, 1, 1994-2005.

4. Pinheiro, A. V.; Han, D. R.; Shih, W. M.; Yan, H. Challenges and Opportunities for Structural DNA Nanotechnology. *Nat. Nanotechnol.* **2011**, 6, 763-772.

5. Douglas, S. M.; Dietz, H.; Liedl, T.; Hoegberg, B.; Graf, F.; Shih, W. M. Self-Assembly of DNA into Nanoscale Three-Dimensional Shapes. *Nature* **2009**, 459, 414-418.

6. Chen, J. H.; Seeman, N. C. Synthesis from DNA of a Molecule with the Connectivity of a Cube. *Nature* **1991**, 350, 631-633.

7. Goodman, R. P.; Schaap, I. A. T.; Tardin, C. F.; Erben, C. M.; Berry, R. M.; Schmidt, C. F.; Turberfield, A. J. Rapid Chiral Assembly of Rigid DNA Building Blocks for Molecular Nanofabrication. *Science* **2005**, 310, 1661-1665.

8. Zhang, Y. W.; Seeman, N. C. Construction of a DNA-Truncated Octahedron. *J. Am. Chem. Soc.* **1994**, 116, 1661-1669.

9. Bhatia, D.; Mehtab, S.; Krishnan, R.; Indi, S. S.; Basu, A.; Krishnan, Y. Icosahedral DNA Nanocapsules by Modular Assembly. *Angew. Chem.-Int. Edit.* **2009**, 48, 4134-4137.

10. Lin, C.; Liu, Y.; Rinker, S.; Yan, H. DNA Tile Based Self-Assembly: Building Complex Nanoarchitectures. *ChemPhysChem* **2006**, 7, 1641-1647.

11. Rothemund, P. W. K.; Papadakis, N.; Winfree, E. Algorithmic Self-Assembly of DNA Sierpinski Triangles. *PLoS Biol.* **2004**, 2, 2041-2053.

12. Yan, H.; Park, S. H.; Finkelstein, G.; Reif, J. H.; LaBean, T. H. DNA-Templated Self-Assembly of Protein Arrays and Highly Conductive Nanowires. *Science* **2003**, 301, 1882-1884.

13. Yan, H.; LaBean, T. H.; Feng, L.; Reif, J. H. Directed Nucleation Assembly of DNA Tile Complexes for Barcode-Patterned Lattices. *Proc. Natl. Acad. Sci. U. S. A.* **2003**, 100, 8103-8108.

14. Wang, T.; Schiffels, D.; Cuesta, S. M.; Fygenson, D. K.; Seeman, N. C. Design and Characterization of 1d Nanotubes and 2d Periodic Arrays Self-Assembled from DNA Multi-Helix Bundles. *J. Am. Chem. Soc.* **2012**, 134, 1606-1616.

15. Kuzuya, A.; Wang, R. S.; Sha, R. J.; Seeman, N. C. Six-Helix and Eight-Helix DNA Nanotubes Assembled from Half-Tubes. *Nano Lett.* **2007**, 7, 1757-1763.

16. Mathieu, F.; Liao, S. P.; Kopatscht, J.; Wang, T.; Mao, C. D.; Seeman, N. C. Six-Helix Bundles Designed from DNA. *Nano Lett.* **2005**, 5, 661-665.

17. Rothemund, P. W. K.; Ekani-Nkodo, A.; Papadakis, N.; Kumar, A.; Fygenson, D. K.; Winfree, E. Design and Characterization of Programmable DNA Nanotubes. *J. Am. Chem. Soc.* **2004**, 126, 16344-16352.

18. Chakraborty, B.; Sha, R.; Seeman, N. C. A DNA-Based Nanomechanical Device with Three Robust States. *Proc. Natl. Acad. Sci. U. S. A.* **2008**, 105, 17245-17249.

19. Seeman, N. C. From Genes to Machines: DNA Nanomechanical Devices. *Trends Biochem. Sci.* **2005**, 30, 119-125.

20. Wickham, S. F. J.; Bath, J.; Katsuda, Y.; Endo, M.; Hidaka, K.; Sugiyama, H.; Turberfield, A. J. A DNA-Based Molecular Motor That Can Navigate a Network of Tracks. *Nat. Nanotechnol.* **2012**, 7, 169-173.

21. Rothemund, P. W. K. Folding DNA to Create Nanoscale Shapes and Patterns. *Nature* **2006**, 440, 297-302.

22. Burns, J. R.; Stulz, E.; Howorka, S. Self-Assembled DNA Nanopores That Span Lipid Bilayers. *Nano Lett.* **2013**, 13, 2351-2356.

23. Langecker, M.; Arnaut, V.; Martin, T. G.; List, J.; Renner, S.; Mayer, M.; Dietz, H.; Simmel, F. C. Synthetic Lipid Membrane Channels Formed by Designed DNA Nanostructures. *Science* **2012**, 338, 932-936.

24. Douglas, S. M.; Marblestone, A. H.; Teerapittayanon, S.; Vazquez, A.; Church, G. M.; Shih, W. M. Rapid Prototyping of 3d DNA-Origami Shapes with Cadnano. *Nucleic Acids Res.* **2009**, gkp436.

25. Lu, X. J.; Olson, W. K. 3dna: A Software Package for the Analysis, Rebuilding and Visualization of Three-Dimensional Nucleic Acid Structures. *Nucleic Acids Res.* **2003**, 31, 5108-5121.

26. Schiffels, D.; Liedl, T.; Fygenson, D. K. Nanoscale Structure and Microscale Stiffness of DNA Nanotubes. *ACS Nano* **2013**, 7, 6700-6710.

27. Kauert, D. J.; Kurth, T.; Liedl, T.; Seidel, R. Direct Mechanical Measurements Reveal the Material Properties of Three-Dimensional DNA Origami. *Nano Lett.* **2011**, 11, 5558-5563.

28. Song, J.; Arbona, J.-M.; Zhang, Z.; Liu, L.; Xie, E.; Elezgaray, J.; Aime, J.-P.; Gothelf, K. V.; Besenbacher, F.; Dong, M. Direct Visualization of Transient Thermal Response of a DNA Origami. *J. Am. Chem. Soc.* **2012**, 134, 9844-9847.

29. Bai, X.-c.; Martin, T. G.; Scheres, S. H.; Dietz, H. Cryo-Em Structure of a 3d DNA-Origami Object. *Proc. Natl. Acad. Sci. U. S. A.* **2012**, 109, 20012-20017.

30. Zheng, J.; Birktoft, J. J.; Chen, Y.; Wang, T.; Sha, R.; Constantinou, P. E.; Ginell, S. L.; Mao, C.; Seeman, N. C. From Molecular to Macroscopic *Via* the Rational Design of a Self-Assembled 3d DNA Crystal. *Nature* **2009**, 461, 74-77.

31. McKinney, S. A.; Déclais, A.-C.; Lilley, D. M.; Ha, T. Structural Dynamics of Individual Holliday Junctions. *Nat. Struct. Mol. Biol.* **2003**, 10, 93-97.

32. Joshi, H.; Dwaraknath, A.; Maiti, P. K. Structure, Stability and Elasticity of DNA Nanotubes. *Phys. Chem. Chem. Phys.* **2015**, 17, 1424-1434.

33. Yoo, J.; Aksimentiev, A. *In Situ* Structure and Dynamics of DNA Origami Determined through Molecular Dynamics Simulations. *Proc. Natl. Acad. Sci. U. S. A.* **2013**, 110, 20099-20104.

34. Yin, P.; Hariadi, R. F.; Sahu, S.; Choi, H. M. T.; Park, S. H.; LaBean, T. H.; Reif, J. H. Programming DNA Tube Circumferences. *Science* **2008**, 321, 824-826.

35. Seeman, N. C. DNA Nicks and Nodes and Nanotechnology. *Nano Lett.* **2001**, 1, 22-26.

36. Holliday, R. Mechanism for Gene Conversion in Fungi. *Genet. Res.* **1964**, 5, 282-&.

37. Zhang, D. Y.; Seelig, G. Dynamic DNA Nanotechnology Using Strand-Displacement Reactions. *Nat. Chem.* **2011**, 3, 103-113.

38. Sa-Ardyen, P.; Vologodskii, A. V.; Seeman, N. C. The Flexibility of DNA Double Crossover Molecules. *Biophys. J.* **2003**, 84, 3829-3837.

39. Fu, T. J.; Seeman, N. C. DNA Double-Crossover Molecules. *Biochemistry* **1993**, 32, 3211-3220.

40. LaBean, T. H.; Yan, H.; Kopatsch, J.; Liu, F. R.; Winfree, E.; Reif, J. H.; Seeman, N. C. Construction, Analysis, Ligation, and Self-Assembly of DNA Triple Crossover Complexes. *J. Am. Chem. Soc.* **2000**, 122, 1848-1860.

41. Shen, Z. Y.; Yan, H.; Wang, T.; Seeman, N. C. Paranemic Crossover DNA: A Generalized Holliday Structure with Applications in Nanotechnology. *J. Am. Chem. Soc.* **2004**, 126, 1666-1674.

42. Sherman, W. B.; Seeman, N. C. Design of Minimally Strained Nucleic Acid Nanotubes. *Biophys. J.* **2006**, 90, 4546-4557.

43. O'Neill, P.; Rothemund, P. W.; Kumar, A.; Fygenson, D. K. Sturdier DNA Nanotubes *Via* Ligation. *Nano Lett.* **2006**, 6, 1379-1383.

44. Liu, D.; Park, S. H.; Reif, J. H.; LaBean, T. H. DNA Nanotubes Self-Assembled from Triple-Crossover Tiles as Templates for Conductive Nanowires. *Proc. Natl. Acad. Sci. U. S. A.* **2004**, 101, 717-722.

45. Lo, P. K.; Karam, P.; Aldaye, F. A.; McLaughlin, C. K.; Hamblin, G. D.; Cosa, G.; Sleiman, H. F. Loading and Selective Release of Cargo in DNA Nanotubes with Longitudinal Variation. *Nat. Chem.* **2010**, 2, 319-328.

46. Yoo, J.; Aksimentiev, A. Molecular Dynamics of Membrane-Spanning DNA Channels: Conductance Mechanism, Electro-Osmotic Transport and Mechanical Gating. *J. Phys. Chem. Lett.* **2015**.

47. Noji, H.; Yasuda, R.; Yoshida, M.; Kinosita, K. Direct Observation of the Rotation of F1-Atpase. *Nature* **1997**, 386, 299-302.

48. Park, S. H.; Barish, R.; Li, H.; Reif, J. H.; Finkelstein, G.; Yan, H.; LaBean, T. H. Three-Helix Bundle DNA Tiles Self-Assemble into 2d Lattice or 1d Templates for Silver Nanowires. *Nano Lett.* **2005**, 5, 693-696.

49. Ouldridge, T. E. DNA Nanotechnology: Understanding and Optimisation through Simulation. *Mol. Phys.* **2015**, 113, 1-15.

50. Falconi, M.; Oteri, F.; Chillemi, G.; Andersen, F. F.; Tordrup, D.; Oliveira, C. L. P.; Pedersen, J. S.; Knudsen, B. R.; Desideri, A. Deciphering the Structural Properties That Confer Stability to a DNA Nanocage. *ACS Nano* **2009**, 3, 1813-1822.

51. Li, C.-Y.; Hemmig, E. A.; Kong, J.; Yoo, J.; Hernández-Ainsa, S.; Keyser, U. F.; Aksimentiev, A. Ionic Conductivity, Structural Deformation, and Programmable Anisotropy of DNA Origami in Electric Field. *ACS Nano* **2015**, 9, 1420-1433.

52. Maingi, V.; Lelimousin, M.; Howorka, S.; Sansom, M. S. P. Gating-Like Motions and Wall Porosity in a DNA Nanopore Scaffold Revealed by Molecular Simulations. *ACS Nano* **2015**, 9, 11209-11217.

53. Arunan, E.; Desiraju, G. R.; Klein, R. A.; Sadlej, J.; Scheiner, S.; Alkorta, I.; Clary, D. C.; Crabtree, R. H.; Dannenberg, J. J.; Hobza, P.; Kjaergaard, H. G.; Legon, A. C.; Mennucci, B.; Nesbitt, D. J. Definition of the Hydrogen Bond (Iupac Recommendations 2011). *Pure Appl. Chem.* **2011**, 83, 1637-1641.

54. Zgarbova, M.; Otyepka, M.; Sponer, J.; Lankas, F.; Jurecka, P. Base Pair Fraying in Molecular Dynamics Simulations of DNA and Rna. *J. Chem. Theory Comput.* **2014**, 10, 3177-3189.

55. Marko, J. F.; Cocco, S. The Micromechanics of DNA. *Phys. World* **2003**, 16, 37-41.

56. Bustamante, C.; Smith, S. B.; Liphardt, J.; Smith, D. Single-Molecule Studies of DNA Mechanics. *Curr. Opin. Struct. Biol.* **2000**, 10, 279-285.

57. Smith, S. B.; Cui, Y. J.; Bustamante, C. Overstretching B-DNA: The Elastic Response of Individual Double-Stranded and Single-Stranded DNA Molecules. *Science* **1996**, 271, 795-799.

58. Marko, J. F.; Siggia, E. D. Stretching DNA. *Macromolecules* **1995**, 28, 8759-8770.

59. Grier, D. G. A Revolution in Optical Manipulation. *Nature* **2003**, 424, 810-816.

60. Cluzel, P.; Lebrun, A.; Heller, C.; Lavery, R.; Viovy, J. L.; Chatenay, D.; Caron, F. DNA: An Extensible Molecule. *Science* **1996**, 271, 792-794.

61. Smith, S. B.; Finzi, L.; Bustamante, C. Direct Mechanical Measurements of the Elasticity of Single DNA-Molecules by Using Magnetic Beads. *Science* **1992**, 258, 1122-1126.

62. Kratky, O.; Porod, G. Rontgenuntersuchung Geloster Fadenmolekule *Recl.: J. R. Neth. Chem. Soc.* **1949**, 68, 1106-1122.

63. Castro, C. E.; Su, H.-J.; Marras, A. E.; Zhou, L.; Johnson, J. Mechanical Design of DNA Nanostructures. *Nanoscale* **2015**, 7, 5913-5921.

64. Mogurampelly, S.; Nandy, B.; Netz, R. R.; Maiti, P. K. Elasticity of DNA and the Effect of Dendrimer Binding. *Eur. Phys. J. E: Soft Matter Biol. Phys.* **2013**, 36.

65. Park, S.; Khalili-Araghi, F.; Tajkhorshid, E.; Schulten, K. Free Energy Calculation from Steered Molecular Dynamics Simulations Using Jarzynski's Equality. *J. Chem. Phys.* **2003**, 119, 3559-3566.

66. Konrad, M. W.; Bolonick, J. I. Molecular Dynamics Simulation of DNA Stretching Is Consistent with the Tension Observed for Extension and Strand Separation and Predicts a Novel Ladder Structure. *J. Am. Chem. Soc.* **1996**, 118, 10989-10994.

67. Ahsan, A.; Rudnick, J.; Bruinsma, R. Elasticity Theory of the B-DNA to S-DNA Transition. *Biophys. J.* **1998**, 74, 132-137.

68. Dhar, A.; Chaudhuri, D. Triple Minima in the Free Energy of Semiflexible Polymers. *Phys. Rev. Lett.* **2002**, 89.

69. Mazur, A. K. Wormlike Chain Theory and Bending of Short DNA. *Phys. Rev. Lett.* **2007**, 98.

70. Samuel, J.; Sinha, S. Elasticity of Semiflexible Polymers. *Phys. Rev. E* **2002**, 66.

71. Garai, A.; Saurabh, S.; Lansac, Y.; Maiti, P. K. DNA Elasticity from Short DNA to Nucleosomal DNA. *J. Phys. Chem. B* **2015**, 119, 11146-11156.

72. Ouldridge, T. E.; Louis, A. A.; Doye, J. P. DNA Nanotweezers Studied with a Coarse-Grained Model of DNA. *Phys. Rev. Lett.* **2010**, 104, 178101.

73. Macke, T. J.; Case, D. A. Modeling Unusual Nucleic Acid Structures. In *Molecular Modeling of Nucleic Acids*, Leontis, N. B.; SantaLucia, J., Eds. 1998; Vol. 682, pp 379-393.

74. D.A. Case, V. B., J.T. Berryman, R.M. Betz, Q. Cai, D.S. Cerutti, T.E. Cheatham, III, T.A. Darden, R.E. Duke, H. Gohlke, *et al*. Amber 14, University of California, San Francisco. **2014**.

75. Humphrey, W.; Dalke, A.; Schulten, K. Vmd: Visual Molecular Dynamics. *J. Mol. Graph.* **1996**, 14, 33-38.

76. Schrödinger. *The Pymol Molecular Graphics System*, Version 1.7.4; www.pymol.org.

77. Case, D. A.; Cheatham, T. E.; Darden, T.; Gohlke, H.; Luo, R.; Merz, K. M.; Onufriev, A.; Simmerling, C.; Wang, B.; Woods, R. J. The Amber Biomolecular Simulation Programs. *J. Comput. Chem.* **2005**, 26, 1668-1688.

78. Darden, T.; York, D.; Pedersen, L. Particle Mesh Ewald: An N· Log (N) Method for Ewald Sums in Large Systems. *J. Chem. Phys.* **1993**, 98, 10089-10092.

79. Cornell, W. D.; Cieplak, P.; Bayly, C. I.; Gould, I. R.; Merz, K. M.; Ferguson, D. M.; Spellmeyer, D. C.; Fox, T.; Caldwell, J. W.; Kollman, P. A. A Second Generation Force Field for the Simulation of Proteins, Nucleic Acids, and Organic Molecules (Vol 117, Pg 5179, 1995). *J. Am. Chem. Soc.* **1996**, 118, 2309-2309.

80. Perez, A.; Marchan, I.; Svozil, D.; Sponer, J.; Cheatham, T. E.; Laughton, C. A.; Orozco, M. Refinenement of the Amber Force Field for Nucleic Acids: Improving the Description of Alpha/Gamma Conformers. *Biophys. J.* **2007**, 92, 3817-3829.

81. Jorgensen, W. L.; Chandrasekhar, J.; Madura, J. D.; Impey, R. W.; Klein, M. L. Comparison of Simple Potential Function for Simulating Liquid Water. *J. Chem. Phys.* **1983**, 79, 926-935.

82. Joung, I. S.; Cheatham, T. E., III. Determination of Alkali and Halide Monovalent Ion Parameters for Use in Explicitly Solvated Biomolecular Simulations. *J. Phys. Chem. B* **2008**, 112, 9020-9041.

83. Le Grand, S.; Götz, A. W.; Walker, R. C. Spfp: Speed without Compromise—a Mixed Precision Model for Gpu Accelerated Molecular Dynamics Simulations. *Comput. Phys. Commun.* **2013**, 184, 374-380.

84. Pearlman, D. A.; Case, D. A.; Caldwell, J. W.; Ross, W. S.; Cheatham, T. E.; Debolt, S.; Ferguson, D.; Seibel, G.; Kollman, P. Amber, a Package of Computer-Programs for Applying Molecular Mechanics, Normal-Mode Analysis, Molecular-Dynamics and Free-Energy Calculations to Simulate the Structural and Energetic Properties of Molecules. *Comput. Phys. Commun.* **1995**, 91, 1-41.

85. Ryckaert, J. P.; Ciccotti, G.; Berendsen, H. J. C. Numerical-Integration of Cartesian Equations of Motion of a System with Constraints - Molecular-Dynamics of N-Alkanes. *J. Comput. Phys.* **1977**, 23, 327-341.

86. Maiti, P. K.; Pascal, T. A.; Vaidehi, N.; Goddard, W. A. The Stability of Seeman Jx DNA Topoisomers of Paranemic Crossover (Px) Molecules as a Function of Crossover Number. *Nucleic Acids Res.* **2004**, 32, 6047-6056.

87. Santosh, M.; Maiti, P. K. Structural Rigidity of Paranemic Crossover and Juxtapose DNA Nanostructures. *Biophys. J.* **2011**, 101, 1393-1402.

88. Maiti, P. K.; Pascal, T. A.; Vaidehi, N.; Goddard, W. A., III. Understanding DNA Based Nanostructures. *J. Nanosci. Nanotechnol.* **2007**, 7, 1712-1720.

89. Maiti, P. K.; Pascal, T. A.; Vaidehi, N.; Heo, J.; Goddard, W. A. Atomic-Level Simulations of Seeman DNA Nanostructures: The Paranemic Crossover in Salt Solution. *Biophys. J.* **2006**, 90, 1463-1479.

90. Evans, E.; Ritchie, K. Dynamic Strength of Molecular Adhesion Bonds. *Biophys. J.* **1997**, 72, 1541-1555.

91. Santosh, M.; Maiti, P. K. Force Induced DNA Melting. *J. Phys.: Condens. Matter* **2009**, 21.

92. Roe, D. R.; Cheatham, T. E., III. Ptraj and Cpptraj: Software for Processing and Analysis of Molecular Dynamics Trajectory Data. *J. Chem. Theory Comput.* **2013**, 9, 3084-3095.

**FIGURES AND TABLES**

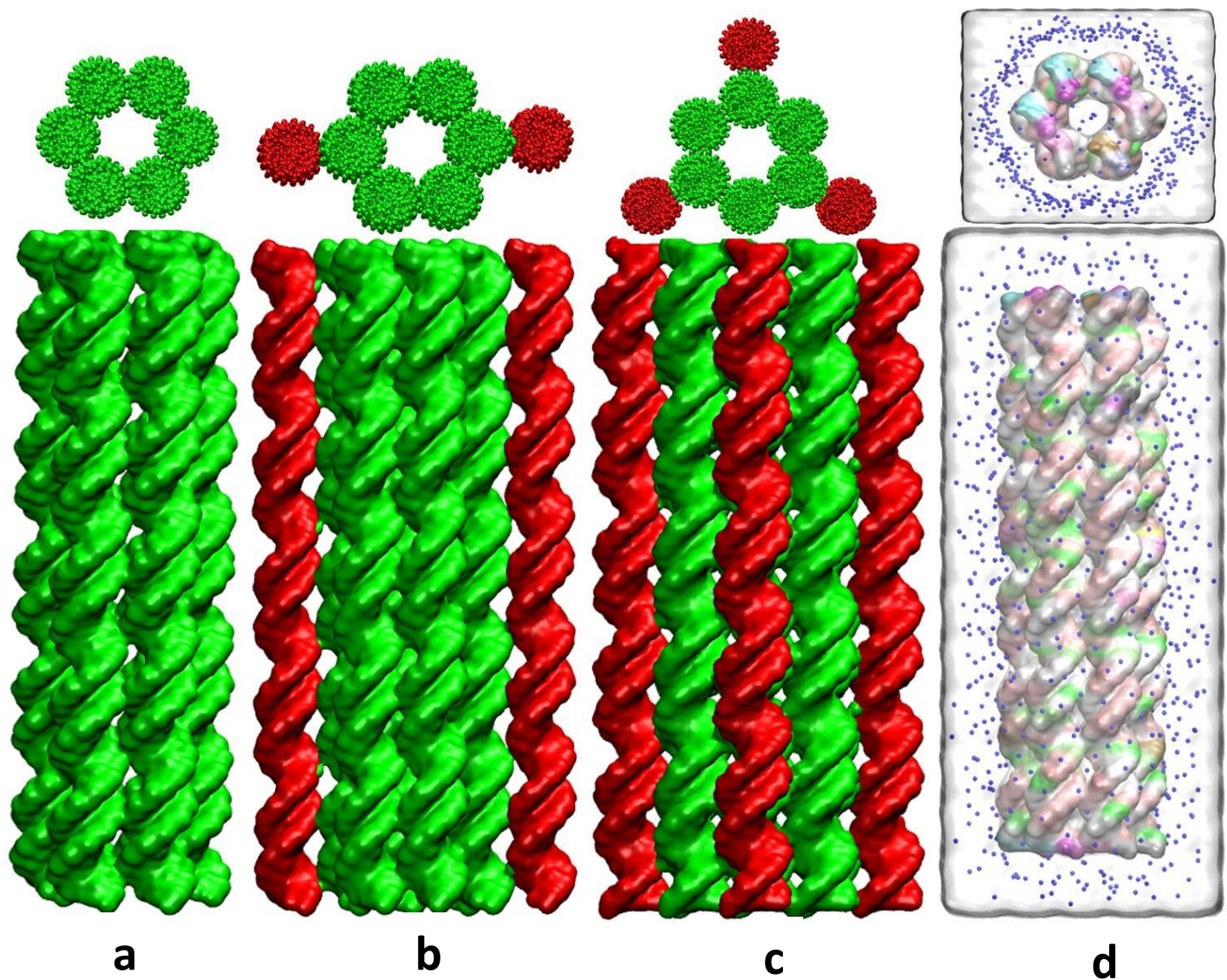

**Figure 1.**

**Starting structures of the various DNTs prior to the MD simulation (a) 6HB, (b) 6HB+2, (c) 6HB+3 (d) 6HB immersed in a TIP3P water box with ions. The external helices in 6HB+2 and 6HB+3, flanking the inner 6HB, have been shown in red. The top panel shows cross-sectional view of the respective structures. The length of each DNT is 57 base pairs.**

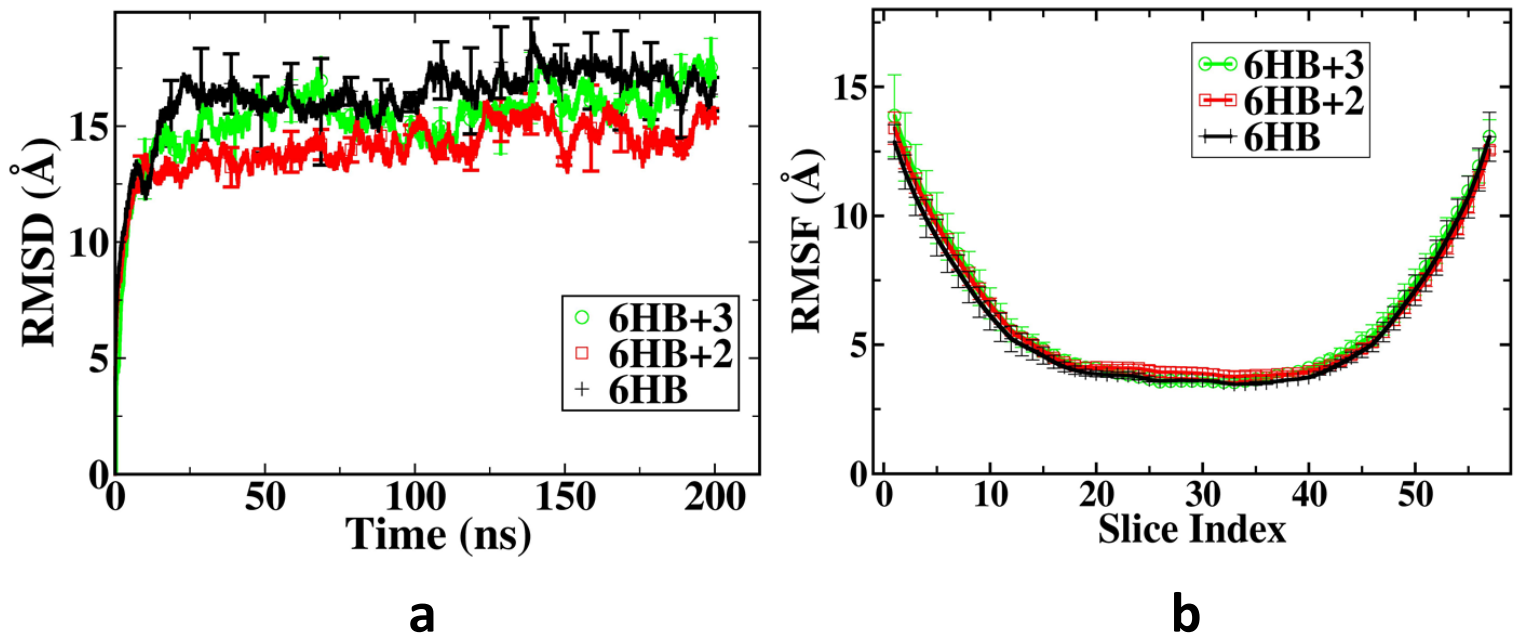

**Figure 2.**

(a) Time Evolution of RMSD for various DNTs with respect to their minimized structures. (b) RMSF fluctuation corresponding to respective slice indices of the DNTs. Both the analyses have been performed for MD trajectories of the hexagonal core of all the structures. The RMSD analysis confirms that the structures are stable during the dynamics whereas the RMSF plot shows that the outer regions of DNTs fluctuate more as compared to the central region. The RMSD and RMSF values have been averaged on three set of statistically independent MD simulation and the standard deviation has also shown on regular intervals respectively.

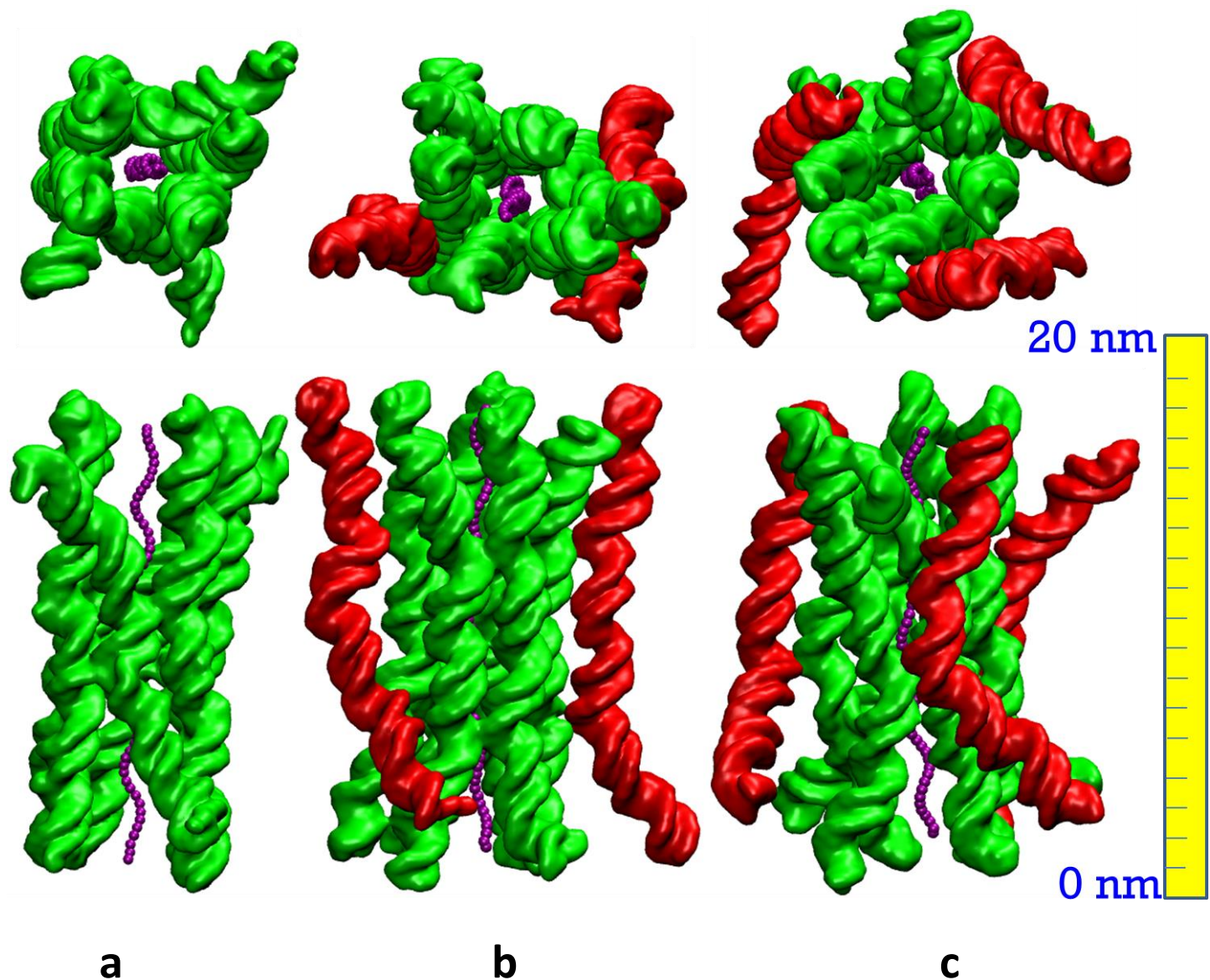

**Figure 3.**

**Instantaneous snapshots of (a) 6HB, (b) 6HB+2 and (c) 6HB+3 DNA nanotubes after 200 ns MD simulation. The top row displays the view of respective snapshot down the vertical axis. The geometrical center of the atoms in each slice has been represented by a bead. The bar at the right side represents the length scale of the figure. The terminals of the DNTs are wide open and distorted with respect to their initial structures.**

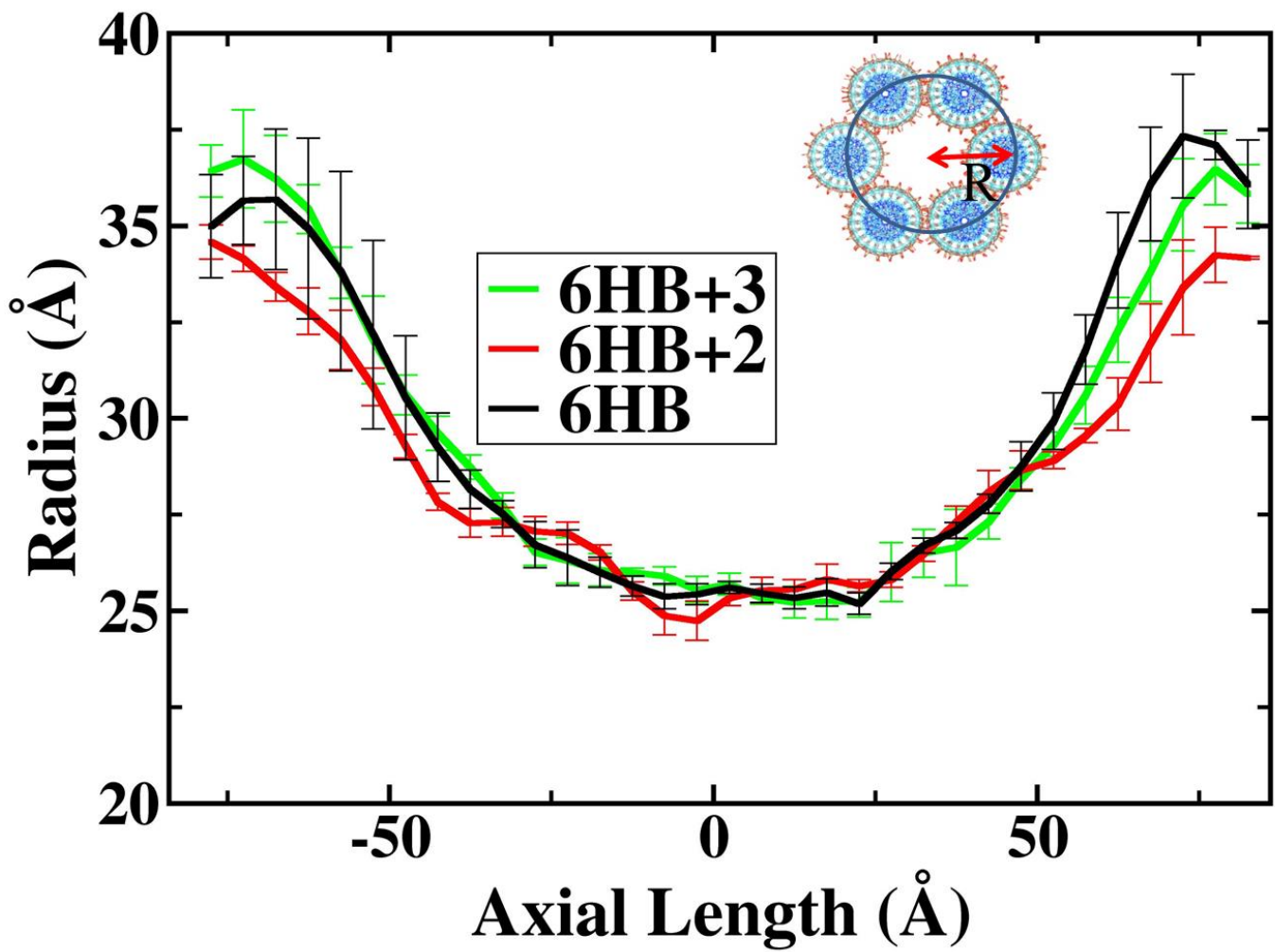

**Figure 4.**

**The radius profiles of DNTs across to their axial length. The central part of the nanotube has a circular cross-section with a radius of ~ 2.5 nm while the ends separate, giving rise to a higher radius ~ 3.5 nm. The notion of the radius also includes the van der Wall radius of DNTs as shown in figure.**

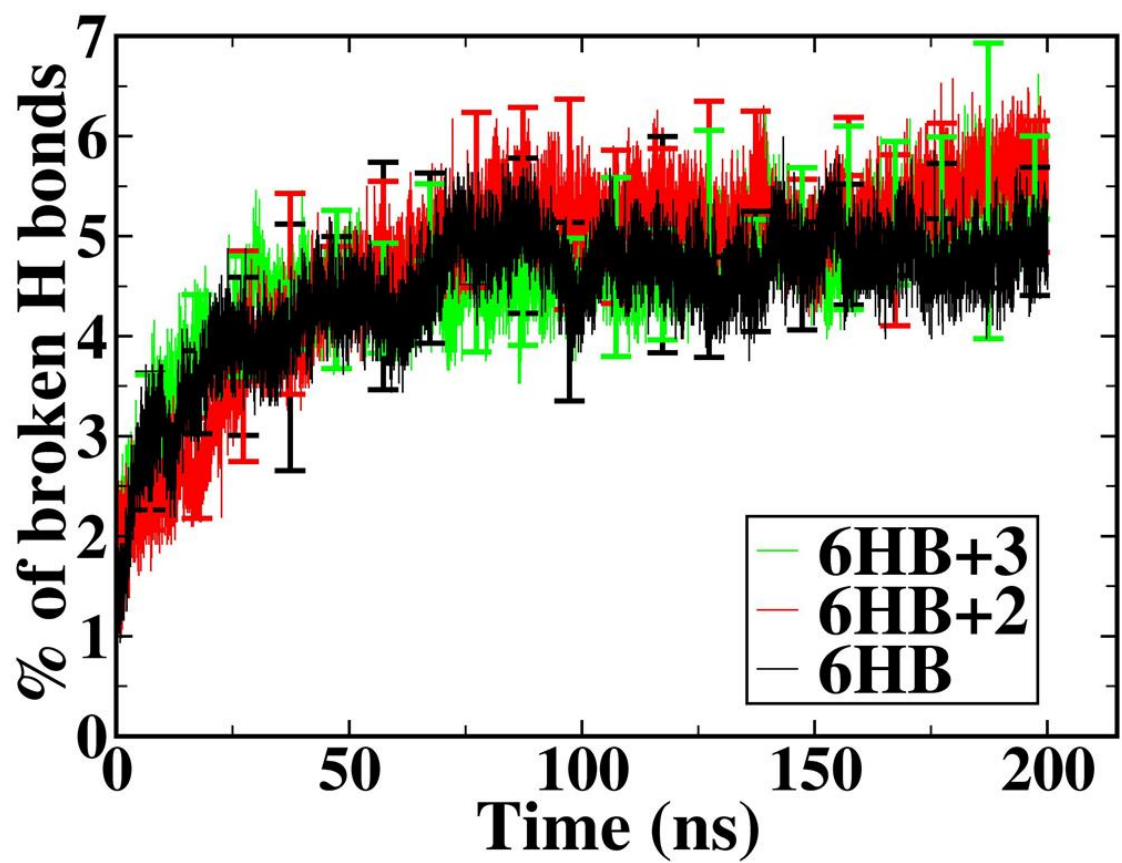

**Figure 5.**

**Time evolution of the percentage of broken hydrogen bonds for the core of the three kinds of DNTs with respect to the simulation time. The percentage of broken hydrogen bonds varies from 0 (initially constructed structure) to ~6 % after 200 ns. The comparison shows that the core of all three structures is equally stable. We see that except for the terminal base pair hydrogen bonds, most of the hydrogen bonds are intact during the simulations. The calculation of broken hydrogen bonds was averaged over three independent MD simulations for each DNT.**

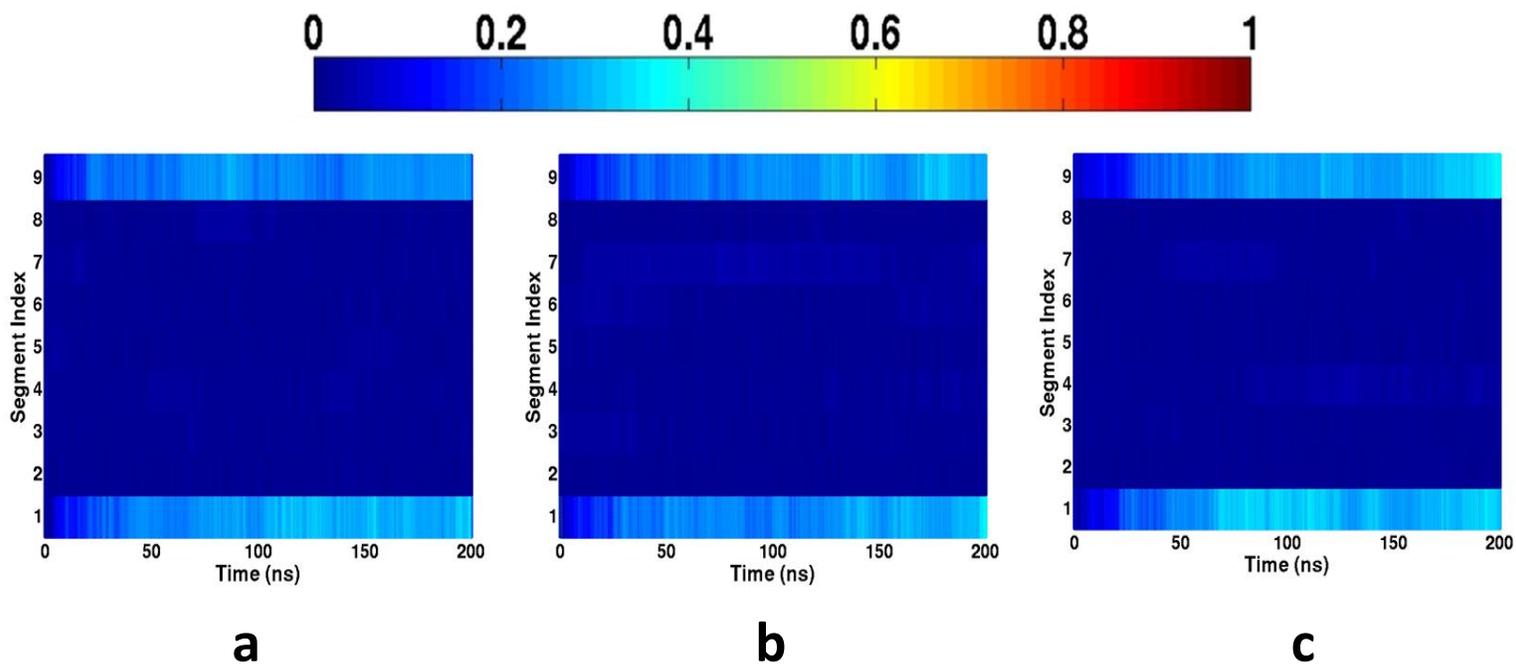

**Figure 6.**

**The dynamical variations of segment-wise fractions of broken hydrogen bond base pairs as a function of the simulation time for the hexagonal cores of (a) 6HB, (b) 6HB+2 and (c) 6HB+3. The blue color stands for all the base pairs intact, while the red for all base pairs broken as shown in the color jet on top of the figure. The terminal segments 1 and 9 show more base pair breakage due to end fraying.**

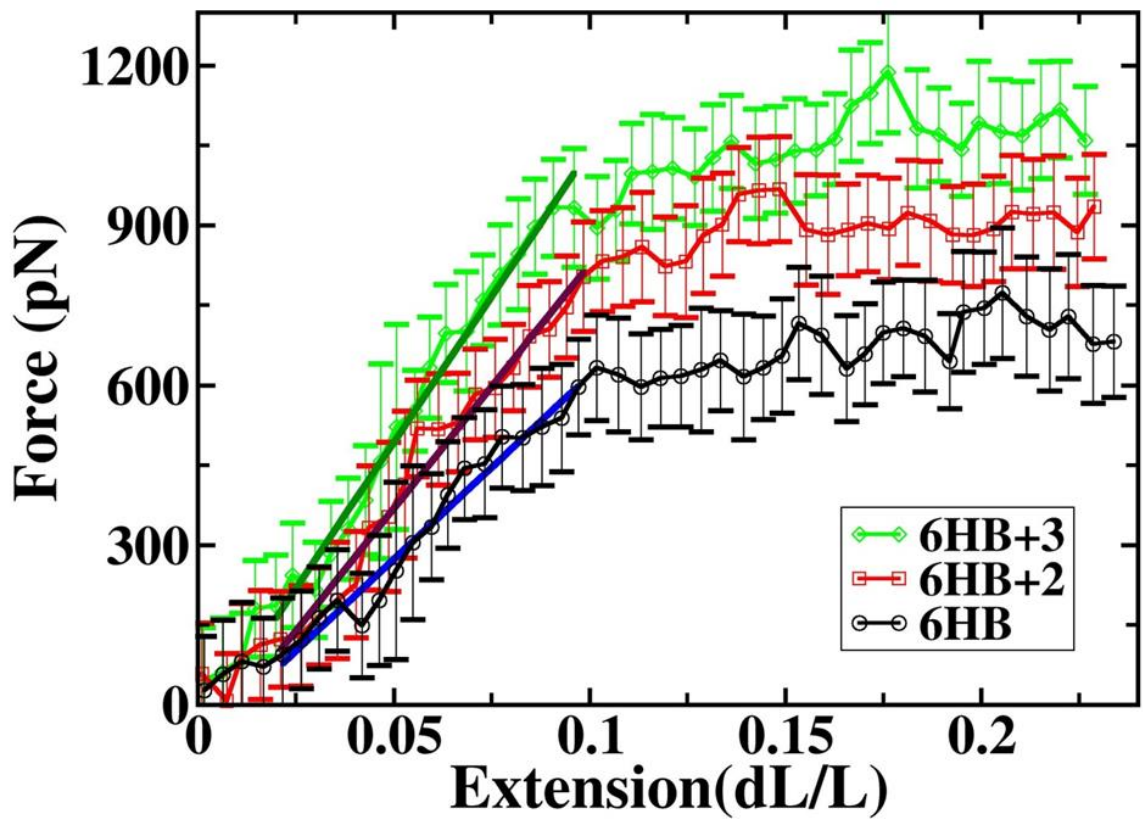

**Figure 7.**

The Force vs extension response of DNTs from the constant velocity SMD simulations. This curve allows us to measure the stretch modulus using Hooke's law. The stretch moduli are computed (Table 4) by fitting the linear region of the force-extension plot as shown in the figure. 6HB+3 DNT shows the highest stiffness followed by 6HB+2 and 6HB respectively.

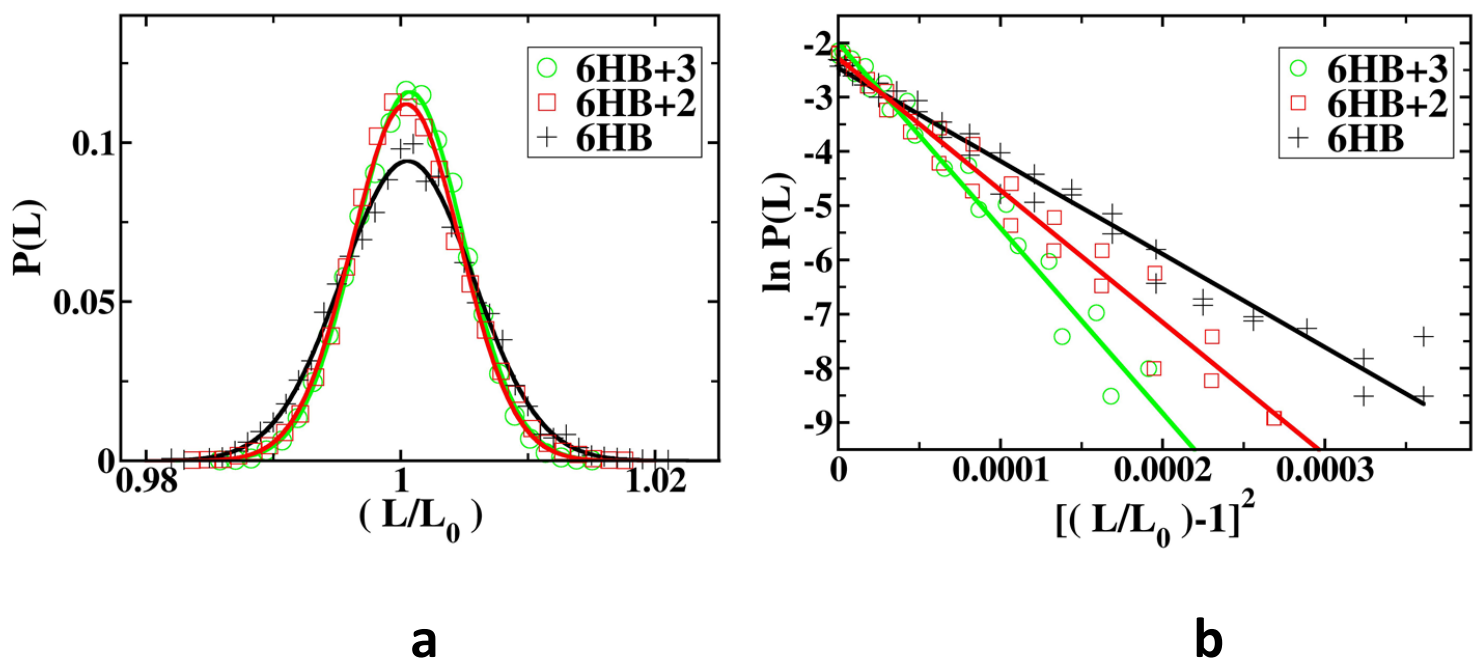

**Figure 8.**

(a) Contour length distributions of DNTs averaged over three independent 200ns MD simulations. The solid line shows the Gaussian fit to Equation 2. (b) The semi log plot of P(L) vs. $[(L/L_0)-1]^2$ using equation 3. The linear fit of this plot has been used to extract the stretch moduli of the DNTs.

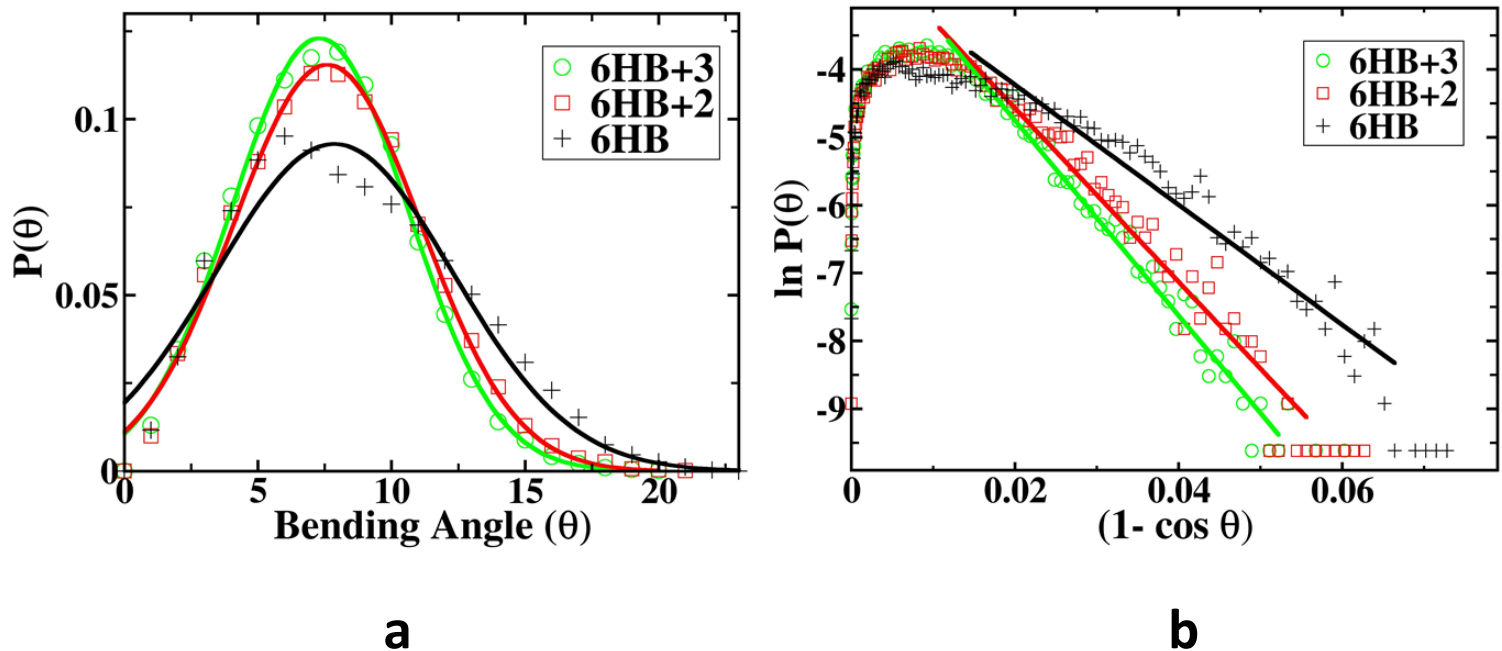

a　　　　　　　　　　　　　　　　　　b

**Figure 9.**

(a) The bend angle distribution of the DNTs, (b) bend angle distribution as a function of 1- cos ($\theta$) in a semi-log plot. The bending angle distribution has been averaged over three independent equilibrium MD simulations for each DNT. The Gaussian nature of the distribution enables us to extract the persistence length using equation 5.

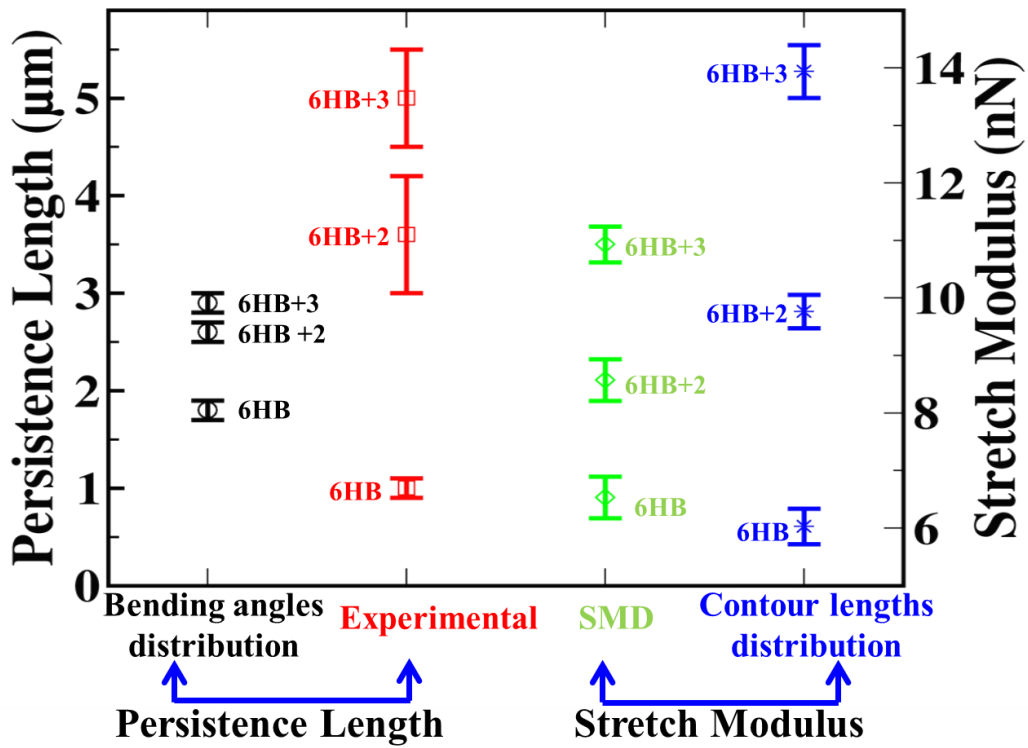

**Figure 10.**

**Comparison of the persistence lengths of 6HB, 6HB+2 and 6HB+3 DNTs obtained using bend angle distribution analysis in equilibrium MD simulations to the values reported by Wang *et al.* in their experiments[14]. The stretch moduli obtained from SMD and contour length distributions from equilibrium MD simulations have also been shown in the other axis for easy comparison.**

# Tables

**Table 1.**

Details of the simulated systems. Three sets of independent simulations were carried out for each system.

| Structure | No. of atoms<br>Total (DNA atoms) | Box dimension<br>[X Y Z] Å | Molarity of Na+ |
|---|---|---|---|
| 6HB | 215175 (21806) | [106 100 242] | 0.424 |
| 6HB+2 | 292950 (29036) | [106 135 242] | 0.429 |
| 6HB+3 | 351403 (32651) | [135 126 242] | 0.403 |

**Table 2.**

**Average helicoidal parameters for the core of DNTs at the end of 200 ns long MD simulation. The standard deviation has been given in the parenthesis.**

**(a) Base Pair Parameters**

| Snapshot Time | Name of the Structure | Shear (Å) | Stretch (Å) | Stagger (Å) | Buckle (°) | Propeller (°) | Opening (°) |
|---|---|---|---|---|---|---|---|
| After 200 ns MD | 6HB | 0.12 (± 1.00) | -0.02 (±1.05) | 0.09 (±0.82) | -1.35 (±16.23) | -14.10 (±14.53) | 2.24 (±11.69) |
| | 6HB+2 | -0.08 (± 1.08) | -0.93 (± 1.52) | 0.15 (±1.32) | -0.01 (± 20.75) | -14.00 (± 17.31) | 2.71 (±17.34) |
| | 6HB+3 | 0.01 (±0.79) | -0.12 (± 1.27) | 0.02 (±1.75) | 0.31 (±23.97) | -12.36 (±17.77) | 4.37 (±19.38) |
| Graphical representation of the parameter | | 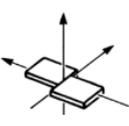 | 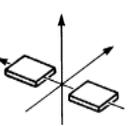 | 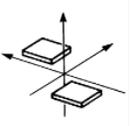 | 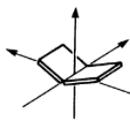 | 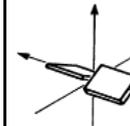 | 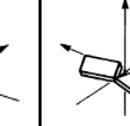 |
| Built | 12-mer B-DNA | 0.00 (±0.08) | 0.12 (±0.03) | 0.0 (±0.0) | 0.0 (±0.05) | 0.03 (±0.01) | 0.0 (±0.08) |

**(b) Base Step Parameters.**

| Snapshots Time | Name of the Structure | Shift (Å) | Slide (Å) | Rise (Å) | Tilt (°) | Roll (°) | Twist (°) |
|---|---|---|---|---|---|---|---|
| After 200 ns MD | 6HB | -0.18 (± 1.87) | - 0.69 (± 2.26) | 3.29 (± 1.33) | 1.78 (± 20.36) | 3.24 (± 9.87) | 30.19 (± 17.65) |
| | 6HB+2 | 0.12 (±1.49) | -0.50 (± 1.04) | 3.36 (± 0.92) | -0.85 (± 12.26) | -2.12 (± 11.66) | 33.56 (± 10.98) |
| | 6HB+3 | - 0.06 (± 2.87) | - 1.00 (± 9.56) | 3.19 (±3.26) | 0.95 (± 24.29) | 2.12 (± 12.25) | 31.14 (± 24.75) |
| Graphical representation of the parameters | | 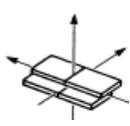 | 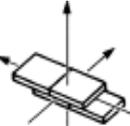 | 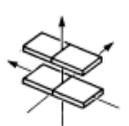 | 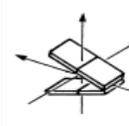 | 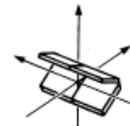 | 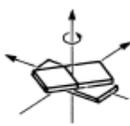 |
| Built | 12-mer B-DNA | 0.00 (±0.01) | -0.27 (±0.03) | 3.37 (±0.01) | 0.03 (±0.27) | 3.20 (±0.21) | 35.71 (±1.28) |

**(c)    Helical Parameters.**

| Snapshots Time | Name of the Structure | X-disp. (Å) | Y-disp. (Å) | Helical Rise (Å) | Inclination (°) | Tip (°) | Helical Twist (°) |
|---|---|---|---|---|---|---|---|
| After 200 ns MD | 6HB | -1.44 (± 2.44) | -0.24 (± 1.88) | 3.24 (± 0.57) | 4.66 (± 14.75) | 1.73 (± 0.09) | 30.53 (± 26.41) |
| | 6HB+2 | -1.63 (± 1.95) | -0.15 (± 1.56) | 3.20 (± 0.46) | 5.73 (± 14.86) | 0.85 (± 0.08) | 35.26 (± 14.75) |
| | 6HB+3 | -1.69 (± 5.35) | 0.11 (± 2.23) | 3.28 (± 0.53) | 5.00 (± 13.22) | 1.03 (± 0.13) | 31.36 (± 31.42) |
| Graphical representation of the parameter | | 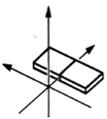 | 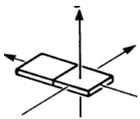 | 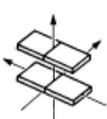 | 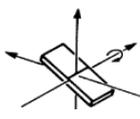 | 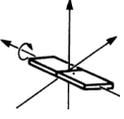 | 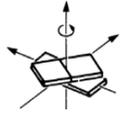 |
| Built | 12-mer B-DNA | 0.02 (±0.03) | 0.00 (±0.01) | 3.38 (±0.01) | -5.21 (±0.39) | 0.05 (±0.05) | 35.85 (±1.27) |

**Table 3.**

**The persistence length of DNTs from the bending angle distribution analysis**

| Structure | Mean Contour Length (Å) | Slope | Persistence Length (μm) |
|---|---|---|---|
| ds-DNA | 43.87 (± 0.72) | 12.10 (± 0.34) | .053 (± .0005) |
| 6HB | 205.0 (± 1.5) | -88.3 (± 2.9) | 1.8 (±.1) |
| 6HB+2 | 205.9 (± 0.9) | -127.7 (± 3.5) | 2.6 (±0.1) |
| 6HB+3 | 205.2 (±0.8) | -143.6 (± 2.7) | 2.9 (±0.1) |

**Table 4:**

**Stretch moduli and persistence lengths of DNTs obtained from the force-extension data and equilibrium contour length distributions. The force extension data has been averaged over two set of SMD simulations. Similarly the contour length distribution has been averaged over three sets of independent equilibrium MD simulations. The persistence length has been calculated assuming that each DNA helix is a rigid cylinder of radius 1 nm. The outer and inner radius of DNTs has been taken to be 2.5 nm and 5.0 nm.**

| | | | Force extension | | Contour length analysis | |
|---|---|---|---|---|---|---|
| **Structure** | **Area ($\pi$ nm$^2$)** | **AMI ($\pi$ nm$^4$)** | **Stretch Moduli (pN)** | **Persistence Length (μm)** | **Stretch Moduli (pN)** | **Persistence Length (μm)** |
| **dS-DNA** | 1 | .25 | **967** ($\pm$58.0) | **.058** $\pm$ 0.003 | **944.45** ($\pm$41.47) | **0.057** **($\pm$.003)** |
| **6HB** | 6 | 20.25 | **6533.3** ($\pm$ 355.5) | **5.3** ($\pm$ 0.3) | **6030.8** ($\pm$ 309.8) | **4.9** ($\pm$0.2) |
| **6HB+2** | 8 | 45.75 | **8753.4** ($\pm$364.6) | **12.1** ($\pm$ 0.5) | **9765.2** ($\pm$291.8) | **13.5** ($\pm$ 0.4) |
| **6HB+3** | 9 | 58.50 | **10932.3** ($\pm$ 317.3) | **17.2** ($\pm$ 0.5) | **13938.6** ($\pm$ 461.2) | **21.9** ($\pm$0.7) |

# SUPPLEMENTARY INFORMATON:

## 1. Sequence and Design of the Atomistic Model of DNA Nanotubes (DNTs)

The initial structures of DNTs have been built using an in-house code written in the NAB[1] built-in of AMBER[2] MD programming suite. The crossovers between the antiparallel strands of dSDNA have been designed carefully so that the inner core of the helices forms a closed hexagon. The sequence of the structures has been supplied to the NAB to code in a specific format as shown below.

ttaaattgatt[1][2]ttaagat[3][4]taaagtt{N}acctcaa[5][6]atatttt[7][8]aatagtc{N}accaaattcaa
aatttaactaa     aattcta     atttcaa     tggagtt     tataaaa     ttatcag     tggtttaagtt

attataatttt     cttcaaattcttgg     agaaagt     aaaacttaacagat     caagttttaa
taatattaaaa[1][2]gaagtttaagaacc [9] [10]tctttca[5][6]ttttgaattgtcta [11] [12] gttcaaaaatt

gtagaaaatgg{N}tatatac[13][14]aatattt[9][10]tataaat{N}aagtcta[15][16]atttttta[11][12]gttaaatcaat
catcttttacc     atatatg     ttataaa     atattta     ttcagat     taaaaat     caatttagtta

atatattataa     aaagcat     gttctag     atcaaat     taagttt     aatcaat     gcttcataagt
tatataatatt[17] [18]tttcgta [13] [14]caagatc{N}tagttta[19] [20]attcaaa[15] [16]ttagtta{N}cgaagtattca

taattcaatct[17] [18]tgatattattatgc[21] [22]ataatgt[19] [20]ataaataaatactt[23] [24]ttgtaatctaa
attaagttaga     actataataatacg     tattaca     tatttatttatgaa     aacattagatt

tctagtatcat     ataagtt     taataat     tcaacat     ctacaaa     aattatt     attataattga
agatcatagta{N}tattcaa[3] [4]attatta[21] [22]agttgta{N}gatgttt[7] [8]ttaataa[23] [24]taatattaact

**The conventions used are as follows –**

- Double helices are separated by blank lines
- Nicks in the backbone are indicated by {N}
- Crossovers connecting corresponding points are indicated by [Crossover_Number]
- Continuous white spaces between letters represent a continuous DNA backbone

The sequence of the inner core of all the structures is similar to the one given above. Figure S1 shows the schematics of the core of the DNTs with strands being numbered.

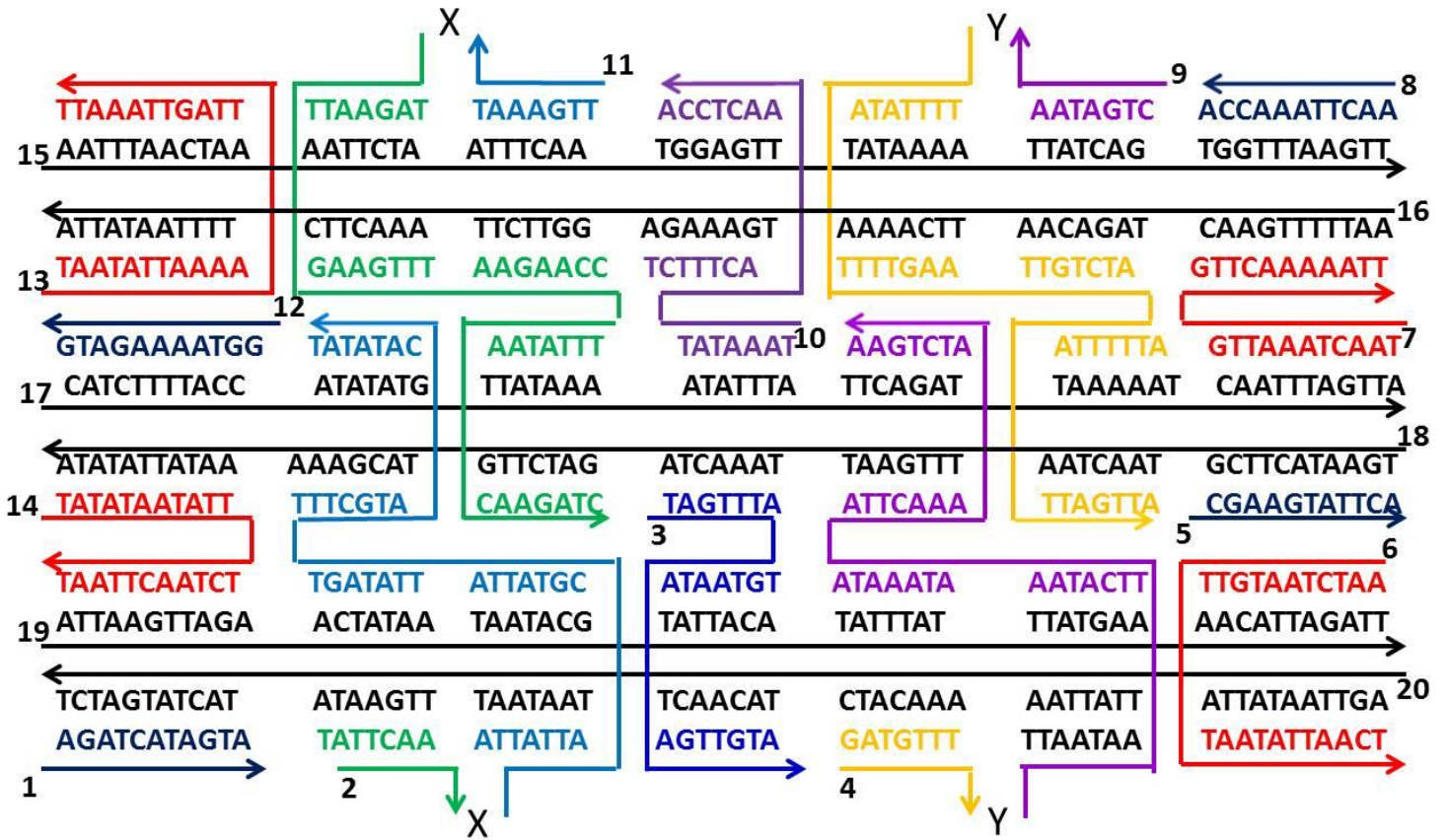

**Figure S1.**

**The sequence of the inner core of DNTs with the numbering of the strands. The strands have been shown with different colors. The arrow represents the 5' end of the oligonucleotide.**

## 2. Geometrical Parameters of DNTs:

Nucleic acids structures can take numerous conformations depending environmental conditions such as water content, pH and ion concentration.[3] These conformations can be fundamentally different from the classical B-DNA. The structure of DNA has been well studied through various spectroscopic and crystallographic techniques and it has been concluded that there exist three sets of translational and rotational parameters, namely base-pair, base step and helical parameters, which can completely describe the secondary structure of DNA.[4] To investigate, the structural aspects of DNTs and their time evolution, we have calculated these geometrical parameters. We have extracted these parameters from the MD simulation snapshots using the CPPTRAJ algorithm.[5] The procedure and terminology used here is similar that adopted in the 3DNA software[6]. Tables 2a-c summarize the values of base pair, base step and helical parameters respectively for all the DNTs at various stages. The respective values of the parameter for NAB-built 12-mer B-DNA has also been shown in the bottom row of the same column in the table. The values of the parameters have been averaged over all the base pairs of the core six DNA double helices and the standard deviations have also been mentioned. The built inner core of all the three structures is identical, and so are the parameters. After the energy minimization, the parameters change slightly. The third row of the table tells that the averaged values of parameters after 200 ns MD simulation vary from the built structures within reasonable limits although there is significant variation in the numbers. All the parameters also compare well with standard 12-mer NAB built B-DNA with the same sequence as the Dickerson dodecamer. A careful assessment of the parameters indicates that most of the deviations are due to distorted base pairs at the termini of each dSDNA. This is to be expected, since the terminal base pairs are known for fraying, thereby causing deviations in the parameters with respect to B-DNA[7]. The geometrical parameters for 6HB, 6HB+2 and 6HB+3 are indistinguishable within the window of the standard deviation. This indicates that the inner hexagonal core of the nanotube is similar in all cases and not perturbed significantly by the addition of outer helices in the case of 6HB+2 and 6HB+3. Overall this analysis reflects that the built geometry of DNTs is preserved after 200 ns the MD simulations. It also validates the construction and designing protocol which has been adopted to build the initial configuration.

**Table S1 (a): The base-pair parameters of DNTs.**

| Snapshots Time | Name of the Structure | Shear (Å) | Stretch (Å) | Stagger (Å) | Buckle (°) | Propeller (°) | Opening (°) |
|---|---|---|---|---|---|---|---|
| Built | 6HB | 0.00 (±0.05) | -0.09 (±0.03) | 0.00 (± 0.00) | 0.00 (±0.03) | - 0.01 (± 0.02) | -6.67 (±0.92) |
| | 6HB+2 | 0.00 (± 0.05) | -0.09 (±0.02) | 0.00 (± 0.00) | 0.00 (±0.03) | -0.01 (± 0.02) | -6.67 (±0.92) |
| | 6HB+3 | 0.00 (±0.05) | -0.09 (±0.02) | 0.00 (± 0.00) | 0.00 (±0.03) | -0.01 (± 0.01) | -6.67 (±0.92) |
| After energy Minimization | 6HB | 0.00 (±0.19) | -0.06 (± 0.07) | -.011 (±0.19) | -0.05 (±3.82) | -3.86 (±2.60) | -5.53 (±1.49) |
| | 6HB+2 | 0.00 (± 0.19) | 0.06 (±0.08) | -0.12 (±0.19) | -0.13 (± 4.08) | -3.83 (±2.78) | -5.56 (± 1.54) |
| | 6HB+3 | 0.01 (±0.19) | -0.05 (±0.07) | -0.13 (± 0.19) | 0.17 (±4.24) | -3.71 (± 2.83) | -5.56 (±1.62) |
| After 200 ns MD | 6HB | 0.12 (± 1.00) | -0.02 (±1.05) | 0.09 (±0.82) | -1.35 (±16.23) | -14.10 (±14.53) | 2.24 (±11.69) |
| | 6HB+2 | -0.08 (± 1.08) | -0.93 (± 1.52) | 0.15 (±1.32) | -0.01 (± 20.75) | -14.00 (± 17.31) | 2.71 (±17.34) |
| | 6HB+3 | 0.01 (±0.79) | -0.12 (± 1.27) | 0.02 (±1.75) | 0.31 (±23.97) | -12.36 (±17.77) | 4.37 (±19.38) |
| Graphical representation of the parameter | | 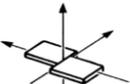 | 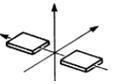 | 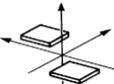 | 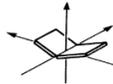 | 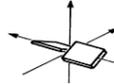 | 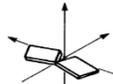 |
| Built | 12-mer B-DNA | 0.00 (±0.08) | 0.12 (±0.03) | 0.0 (±0.0) | 0.0 (±0.05) | 0.03 (±0.01) | 0.0 (±0.08) |

**Table S1 (b): The base-step parameters of DNTs.**

| Snapshots Time | Name of the Structure | Shift (Å) | Slide (Å) | Rise (Å) | Tilt (°) | Roll (°) | Twist (°) |
|---|---|---|---|---|---|---|---|
| Built | 6HB | 0.22 (± 2.68) | -0.27 (± 0.53) | 3.37 (± 0.04) | 0.03 (±0.38) | -3.23 (±0.18) | 34.17 (± 0.63) |
| | 6HB+2 | 0.00 (± 0.03) | -0.31 (± 0.12) | 3.37 (± 0.00) | 0.03 (± 0.38) | -3.22 (± 0.18) | 34.14 (± 0.68) |
| | 6HB+3 | 0.10 (±1.91) | -0.28 (± 0.45) | 3.37 (± 0.03) | 0.00 (± 0.38) | -3.22 (± 0.18) | 34.12 (± 0.62) |
| After energy Minimization | 6HB | 0.22 (± 2.71) | -0.15 (± 0.57) | 3.38 (± 0.12) | 0.04 (± 1.59) | -2.66 (± 1.29) | 34.21 (± 2.17) |
| | 6HB+2 | 0.00 (±0.12) | -0.19 (± 0.12) | 3.38 (± 0.12) | 0.03 (± 1.62) | -2.62 (± 1.34) | 34.24 (± 2.20) |
| | 6HB+3 | 0.11 (±1.93) | -0.16 (± 0.47) | 3.38 (±0.13) | -0.01 (±1.78) | -2.57 (± 1.38) | 34.16 (± 2.08) |
| After 200 ns MD | 6HB | -0.18 (± 1.87) | - 0.69 (± 2.26) | 3.29 (± 1.33) | 1.78 (± 20.36) | 3.24 (± 9.87) | 30.19 (± 17.65) |
| | 6HB+2 | 0.12 (±1.49) | -0.50 (± 1.04) | 3.36 (± 0.92) | -0.85 (± 12.26) | -2.12 (± 11.66) | 33.56 (± 10.98) |
| | 6HB+3 | - 0.06 (± 2.87) | - 1.00 (± 9.56) | 3.19 (±3.26) | 0.95 (± 24.29) | 2.12 (± 12.25) | 31.14 (± 24.75) |
| Graphical representation of the parameters | | 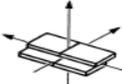 | 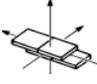 | 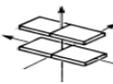 | 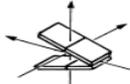 | 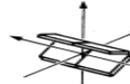 | 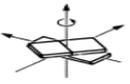 |
| Built | 12-mer B-DNA | **0.00 (±0.01)** | **-0.27 (±0.03)** | **3.37 (±0.01)** | **0.03 (±0.27)** | **3.20 (±0.21)** | **35.71 (±1.28)** |

**Table S1 (c): The helical parameters of DNTs.**

| Snapshots Time | Name of the Structure | X-disp. (Å) | Y-disp. (Å) | Helical Rise (Å) | Inclination (°) | Tip (°) | Helical Twist (°) |
|---|---|---|---|---|---|---|---|
| Built | 6HB | 0.06 (± 0.81) | -0.33 (± 4.09) | 3.38 (±0.13) | -5.48 (± 0.32) | -0.04 (± 0.05) | 34.31 (± 0.63) |
| | 6HB+2 | 0.00 (± 0.03) | 0.00 (±0.02) | 3.38 (± 0.00) | -5.48 (± 0.32) | -0.05 (± 0.00) | 34.29 (± 0.68) |
| | 6HB+3 | 0.04 (± 0.72) | -0.17 (± 3.06) | 3.38 (±0.04) | -5.48 (± 0.32) | 0.00 (± 0.05) | 34.27 (± 0.62) |
| After energy Minimization | 6HB | 0.17 (± 0.84) | -0.31 (± 4.02) | 3.38 (± 0.16) | -4.59 (± 2.35) | -0.07 (± 0.05) | 34.37 (± 2.11) |
| | 6HB+2 | 0.11 (± 0.19) | 0.00 (± 0.37) | 3.38 (± 0.12) | -4.52 (± 2.44) | -0.06 (± 0.02) | 34.40 (± 2.14) |
| | 6HB+3 | 0.15 (± 0.71) | -0.17 (± 2.90) | 3.38 (±0.12) | -4.44 (± 2.51) | 0.02 (± 0.05) | 34.33 (± 2.02) |
| After 200 ns MD | 6HB | -1.44 (± 2.44) | -0.24 (± 1.88) | 3.24 (± 0.57) | 4.66 (± 14.75) | 1.73 (± 0.09) | 30.53 (± 26.41) |
| | 6HB+2 | -1.63 (± 1.95) | -0.15 (± 1.56) | 3.20 (± 0.46) | 5.73 (± 14.86) | 0.85 (± 0.08) | 35.26 (± 14.75) |
| | 6HB+3 | -1.69 (± 5.35) | 0.11 (± 2.23) | 3.28 (± 0.53) | 5.00 (± 13.22) | 1.03 (± 0.13) | 31.36 (± 31.42) |
| Graphical representation of the parameter | | 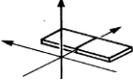 | 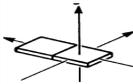 | 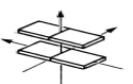 | 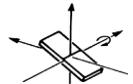 | 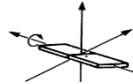 | 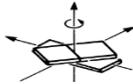 |
| Built | 12-mer B-DNA | **0.02 (±0.03)** | **0.00 (±0.01)** | **3.38 (±0.01)** | **-5.21 (±0.39)** | **0.05 (±0.05)** | **35.85 (±1.27)** |

## 3. The RMSD and RMSF Including the Pillars.

While comparing the RMSD and RMSF in figure 2 in the main manuscript, we excluded the outer helices and compared only the cores of all the structures. In Figure S2, we compare the RMSD and RMSF of the complete structures (core and external helices) of 6HB+2 and 6HB+3. The RMSD and RMSF values of 6HB+3 are higher than 6HB+2.

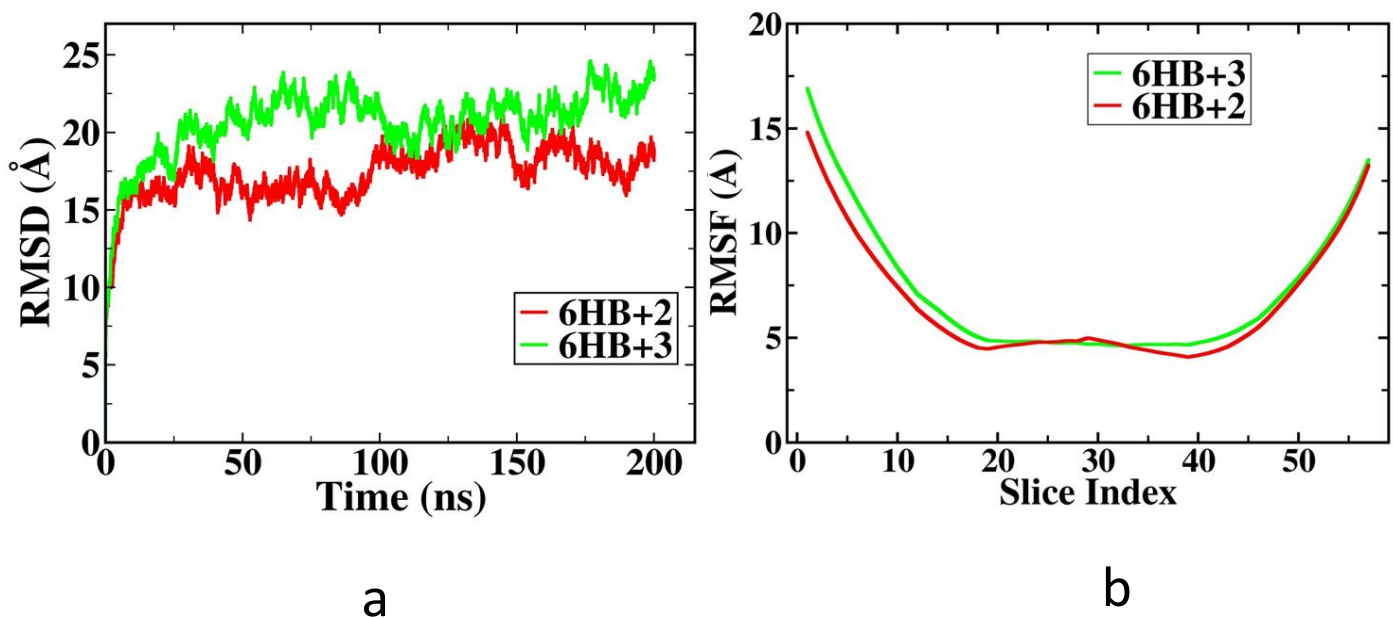

**Figure S2.**

**(a )The RMSD  (b) the RMSF comparison between complete 6HB+2 and 6HB+3 DNTs.**

## 4. The Axis of DNTs and Radius Calculation

The axis of the DNT has been found by dividing the nanotube into small sections along its length. The geometrical center of each section is used to calculate the average radius of that particular section. This radius is plotted against the length of nanotube in Figure 4 in the main manuscript. The axis formed by joining these geometrical centers with the DNT has been show in the Figure S4.

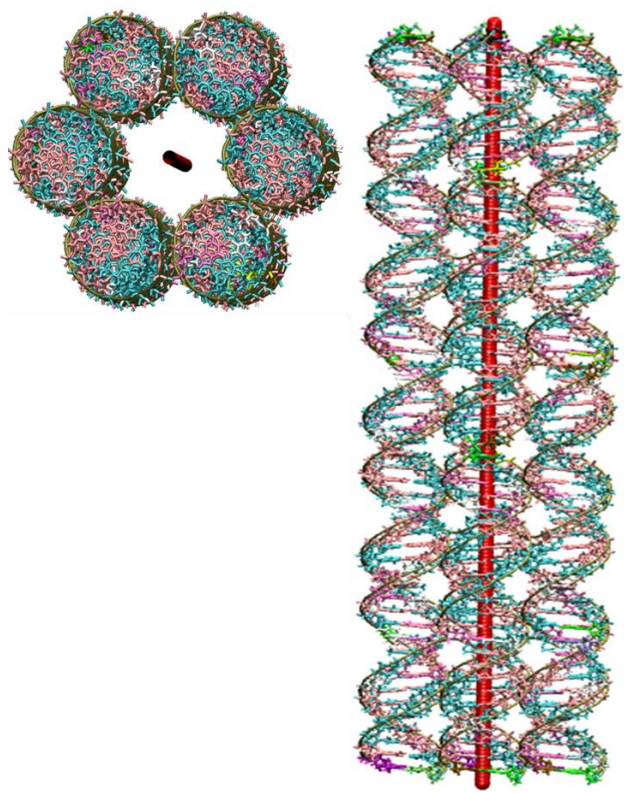

**Figure S3.**

**Representation of a DNT along with its axis; a cross-sectional view is shown on the top left side. The radius at an axial separation of 1 Å has been calculated and shown in various plots in the current study.**

## 5. Division of DNTs into "Slices" and "Segments" for Various Analyses.

We use the notion "Slice" and "Segment" in many analyses presented in this study. Each DNTs is composed of 57 bp per helical domain. DNTs have been divided in to 57 slices containing 1 base pair from each double helix. Average of the RMSF has been calculated for the all atoms of DNA in each slice and plotted with respect to slice index in Figure 2(b). The DNT has been further dived into 9 symmetrical segments where each segment contains 7 slices, except terminal sections 1 and 9, which have four slices each. The two blue lines in figure S4 represent the limits of the sections and the red dots represent their geometrical center. This definition has been used in the analysis of broken base pairs and for computing the bend angle.

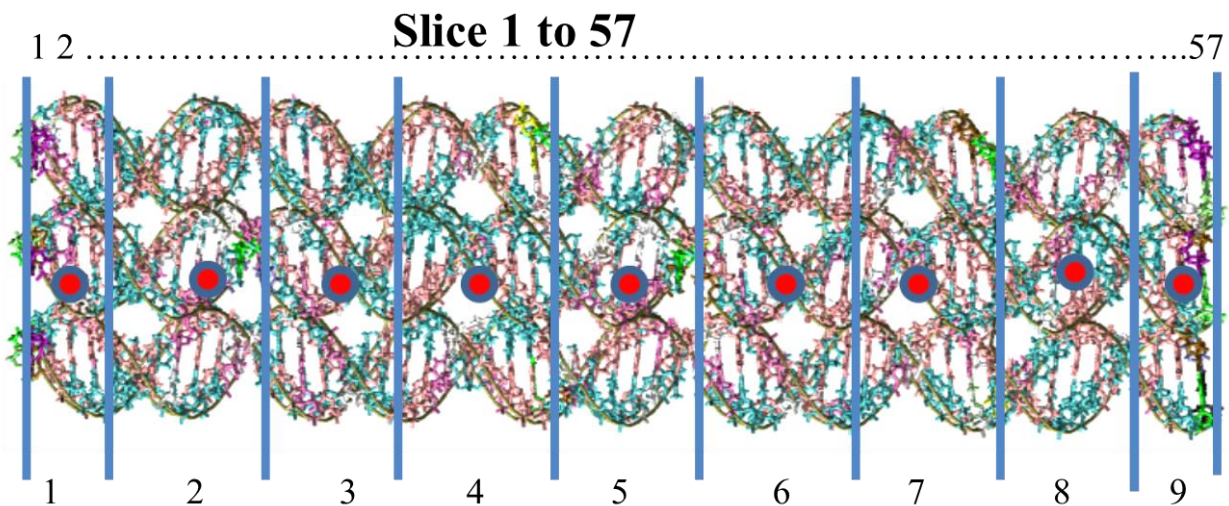

**Figure S4.**

**The division of DNT into slices and segments for various analyses. The slices are represented using small black dots above the image and are numbered from 1 to 57. The blue lines are the boundaries of the segments and the red dot represents the center of each segment.**

## 6. The Broken Base Pair and Hydrogen Bond Analysis for Full 6HB+2 and 6HB+3 DNTs and Effect of Pillars.

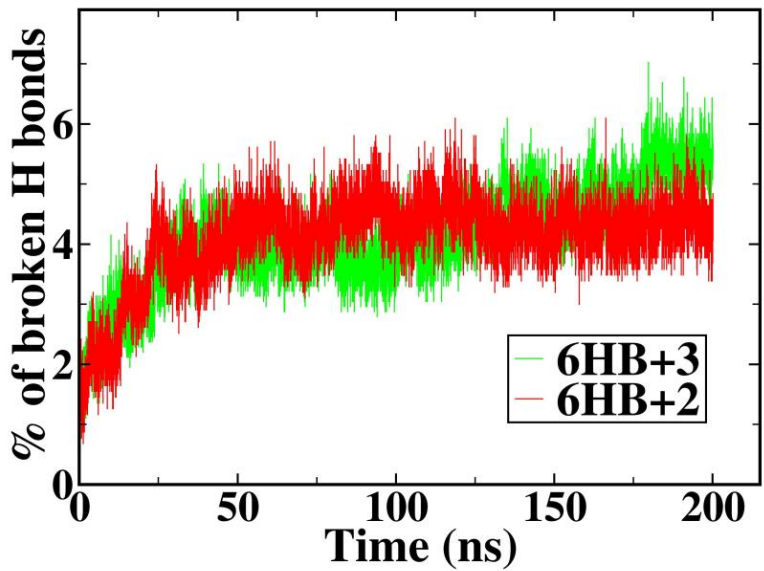

a

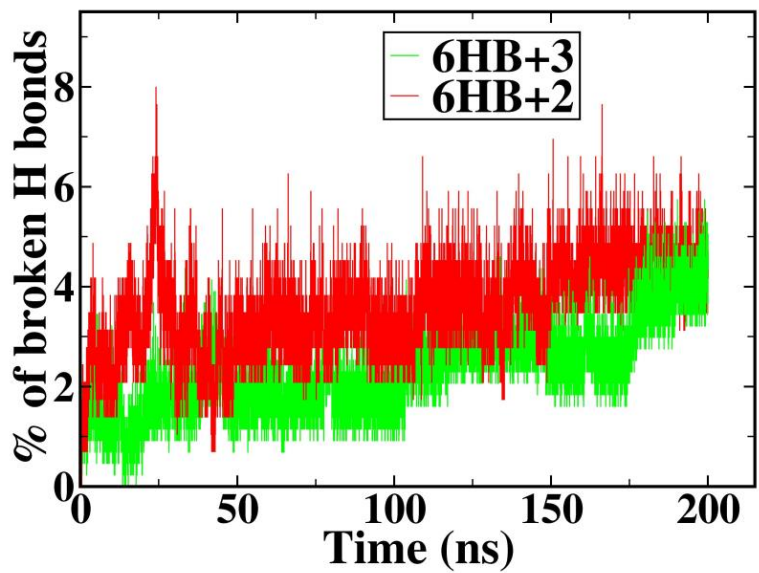

b

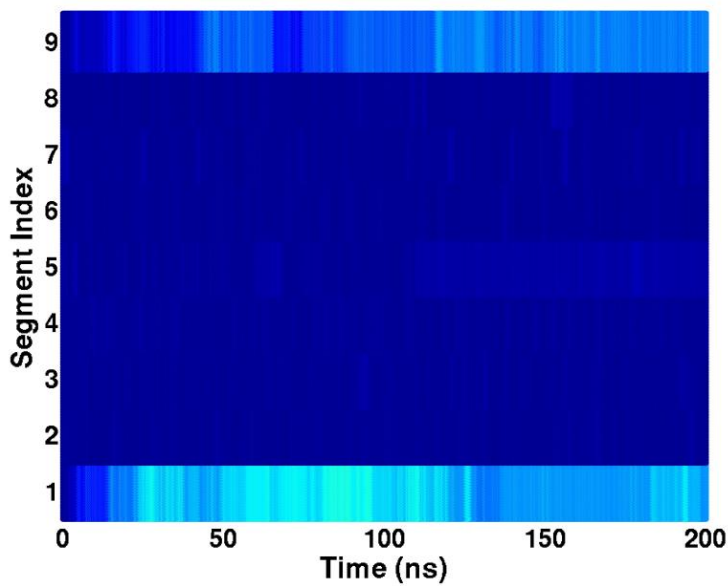

c

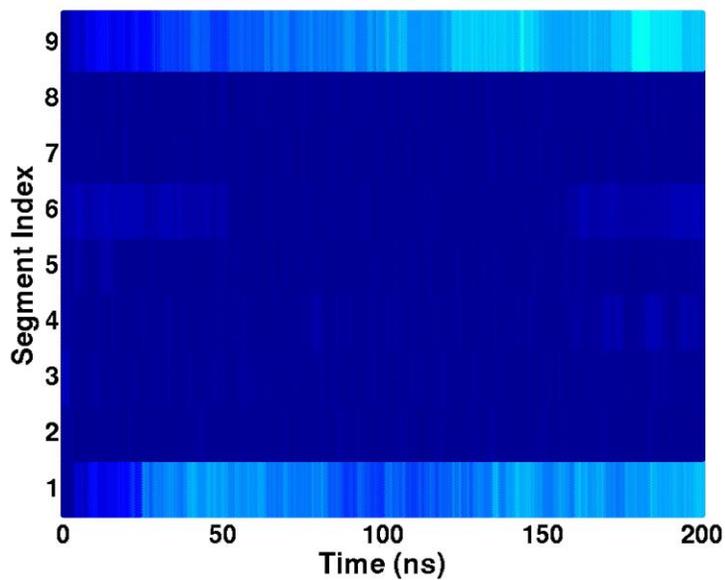

d

**Figure S5.**

**The percentage of broken hydrogen bonds between the base pairs for, (a) full 6HB+2 and 6HB+3 DNTs and (b) the external helices of 6HB+2 and 6HB+3. (c) and (d) represent the dynamics of the fraction of broken base pairs with simulation time for 6HB+2 and 6HB+3 DNTs respectively.**

A base pair is considered as being broken if any one of its constituent hydrogen bonds are broken. For the purposes of comparison, the helices external to the hexagonal core have not been considered in Figures 5 and 6. Here in figure S5 (a) we present the fraction of broken hydrogen bonds between the base pairs for the full DNTs. From Figure S5 (b) we see that the external helices of 6HB+2 have more broken hydrogen bonds than those of the 6HB+3 structure. Figures S5 (c) and (d) show the variation of the segment wise fraction of broken base pairs with respect to the simulation time for 6HB+2 and 6HB+3, respectively. The analysis tells us that the hydrogen bonds in external helices are intact, similar to the core, during the simulation. However, the helical geometry has been significantly distorted as seen in the snapshots of the system after 200 ns.

## 7. The Radius Profile of the Pillars of 6HB+2 and 6HB+3

The radius of the DNT outer region with respect to the axial length has also been analyzed using a protocol similar to that used for the core region (Figure 4 in main manuscript). The last 10 ns of the trajectories have been used for computation of the radius. Figure S7 shows the radius profile of outer region and the inner core with the error bars. Due to the large fluctuations in the outer helices, defining a radius for them is very difficult. The fluctuations lead to a zigzag radius profile, as shown below with an average value of about 5 n. The radius of the outer cores of both the nanotubes can be approximated to be 5 nm. This radius also includes the van der Waals radii of DNA double helix atoms.

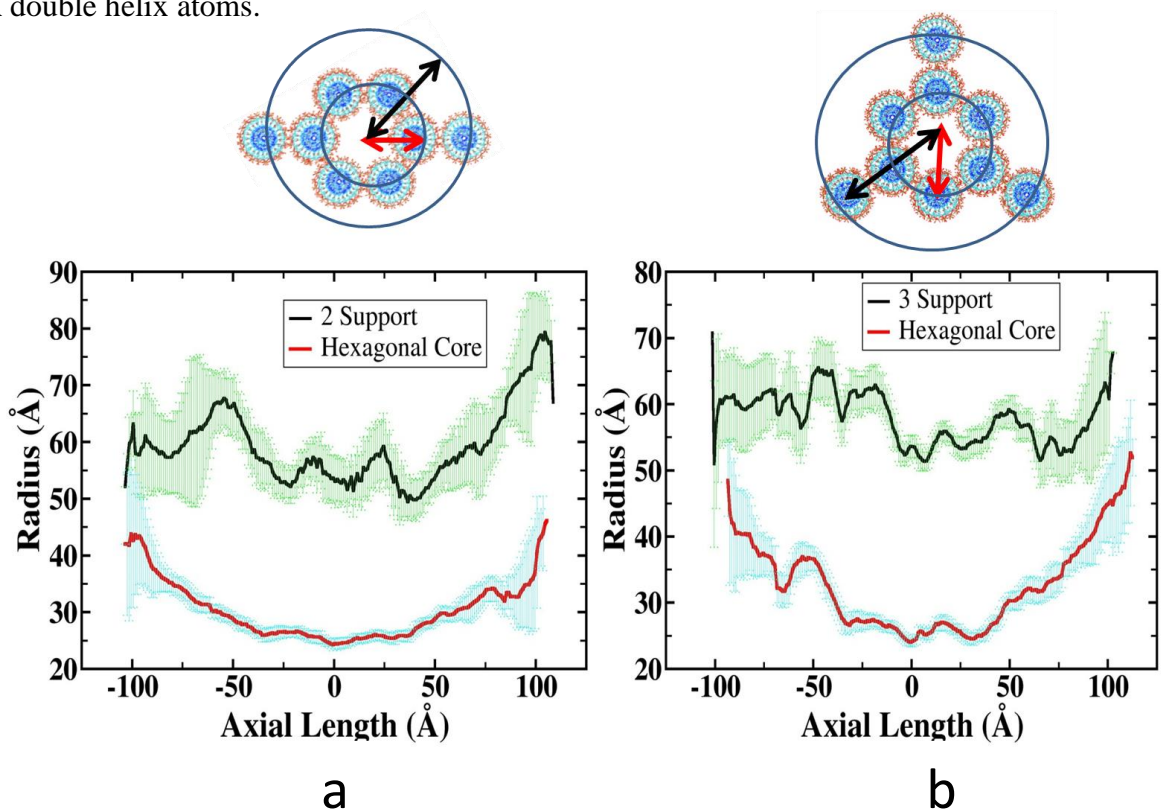

**Figure S6.**

**The comparison between the radii of the core helices and the outer helices of (a) 6HB+2 (b) 6HB+3 DNTs. The radius of the central region of the DNTs is approximately equal to 2.5 and 5 nm for the core and outer helices respectively. The pictures on top show the notion of inner and outer radii.**

## 8: The Bending Axis of DNTs and Estimation of Area Moment of Inertia

The persistence lengths of DNTs, $L_p$, has been calculated using the following relationship

$$L_p = \frac{EI}{k_B T}$$

Where, $E$ is the Young's modulus of DNT, $I$ is the area moment of inertia (AMI), $k_B$ is the Boltzmann constant and $T$ is the absolute temperature.

The AMI, $I$, of a plane with respect to its bending axis is defined as

$$I = \int d^2 dA$$

Where $d$ is the distance of the area element $dA$ from the reference bending axis. In order to calculate the AMI of DNTs, we apply the parallel axis theorem of moment of inertia as follows

$$I_{DNANanotube} = \sum_n (I_0) + Ad^2$$

Here, $n$ is the number of dSDNA components in the respective DNT, $I_0$ is the AMI of the individual dSDNA with respect to the axis passing through its geometrical center of area and perpendicular to its helical axis, $d$ is the perpendicular distance between the aforementioned axis and the bending axis and $A$ is the cross-sectional area of dSDNA.

The AMI for DNTs has been averaged over two perpendicular bending axes as shown in the figure S8. In the above equations, we invoke our radius analysis to calculate the perpendicular distance $d$. The radius of the inner and outer helices has been approximated to be 2.5 and 5.0 nm respectively as shown in Figure 4 of the main manuscript and Figure S6 in the Supplementary Information. Assuming the dSDNA to be a cylindrical object of radius 1 nm, its area $A$, is taken to be $\pi$ nm$^2$. From this assumption, we get the AMI of dSDNA, $I_0$, to be $0.25\pi$ nm$^4$. The bending axis has been illustrated in following Figure S7.

For example, the AMI of 6HB (radius R) with respect to the bending axis 1 is calculated as follows

$$I_{bending\_axis\_1} = \sum_6 [I_0] + 4(r^2 A) + 2(R^2 A)$$

$$= 6I_0 + 16I_0 + 8\frac{R^2}{r^2}I_0$$

$$= I_0(22 + 8\frac{R^2}{r^2})$$

$$I_{bending\_axis\_1} = \sum_6 [I_0] + 4(d_1^2 A) + 2(d_2^2 A)$$

Here, $d_1$ and $d_2$ turns out to be 1 nm and 2.5 nm and putting the values of $I_0$ and A, we get

$$I_{bending\_axis\_1} = (6I_0 + 4A + 12.50A) nm^4$$

$$= 18\pi\, nm^4$$

Similarly, we also calculated the AMI with respect to another bending axis perpendicular to the current one and arithmetically average them. Using the same model, we calculate the AMI of 6HB+2 and 6HB+3 DNTs.

$$I_{bending\_axis\_2} = \sum_6 [I_0] + 4(R^2 - r^2)A)$$

$$= 6I_0 + 16(\frac{R^2}{r^2} - 1)I_0$$

$$= I_0(16\frac{R^2}{r^2} - 10)$$

The inner and outer radii of 6HB+2 and 6HB+3, $R_1$ and $R_2$ respectively, have been shown schematically, in Figure S7.

The AMI for 6HB+2 with respect to the bending axis 1 can be given as

$$I_{bending\_axis\_1} = \sum_6 [I_0] + 4(r^2 A) + 2(R_1^2 A) + 2(I_0 + R_2^2 A)$$

$$= 8I_0 + 16I_0 + 8\frac{R_1^2}{r^2}I_0 + 8\frac{R_2^2}{r^2}I_0$$

$$= I_0[24 + 8\frac{(R_1^2 + R_2^2)}{r^2}]$$

The AMI for 6HB+2 with respect to bending axis 2 can be given as

$$I_{bending\_axis\_2} = \sum_6 [I_0] + 4(R_1^2 - r^2)A + 2I_0$$

$$= 8I_0 + 16(\frac{R_1^2}{r^2} - 1)I_0$$

$$= I_0(16\frac{R_1^2}{r^2} - 8)$$

The AMI of 6HB+3 with respect to the bending axis 1 can be given as

$$I_{bending\_axis\_1} = \sum_6 [I_0] + 4(r^2 A) + 2(R_1^2 A) + I_0 + R_2^2 A + 2(I_0 + R_1^2 A)$$

$$= 9I_0 + 16I_0 + 16\frac{R_1^2}{r^2}I_0 + 4\frac{R_2^2}{r^2}I_0$$

$$= I_0[25 + 4\frac{(4R_1^2 + R_2^2)}{r^2}]$$

The AMI of 6HB+3 with respect to the bending axis 2 is calculated as following

$$I_{bending\_axis\_2} = \sum_6 [I_0] + 4(R_1^2 - r^2)A) + I_0 + 2[I_0 + (R_2^2 - R_1^2)A]$$

$$= 9I_0 + 16(\frac{R_1^2}{r^2} - 1)I_0 + \frac{8}{r^2}(R_2^2 - R_1^2)I_0$$

$$= I_0[8\frac{(R_1^2 + R_2^2)}{r^2} - 7]$$

The values of the AMI for all DNTs from above equations have been shown in the following table S2 along with their respective bending axis.

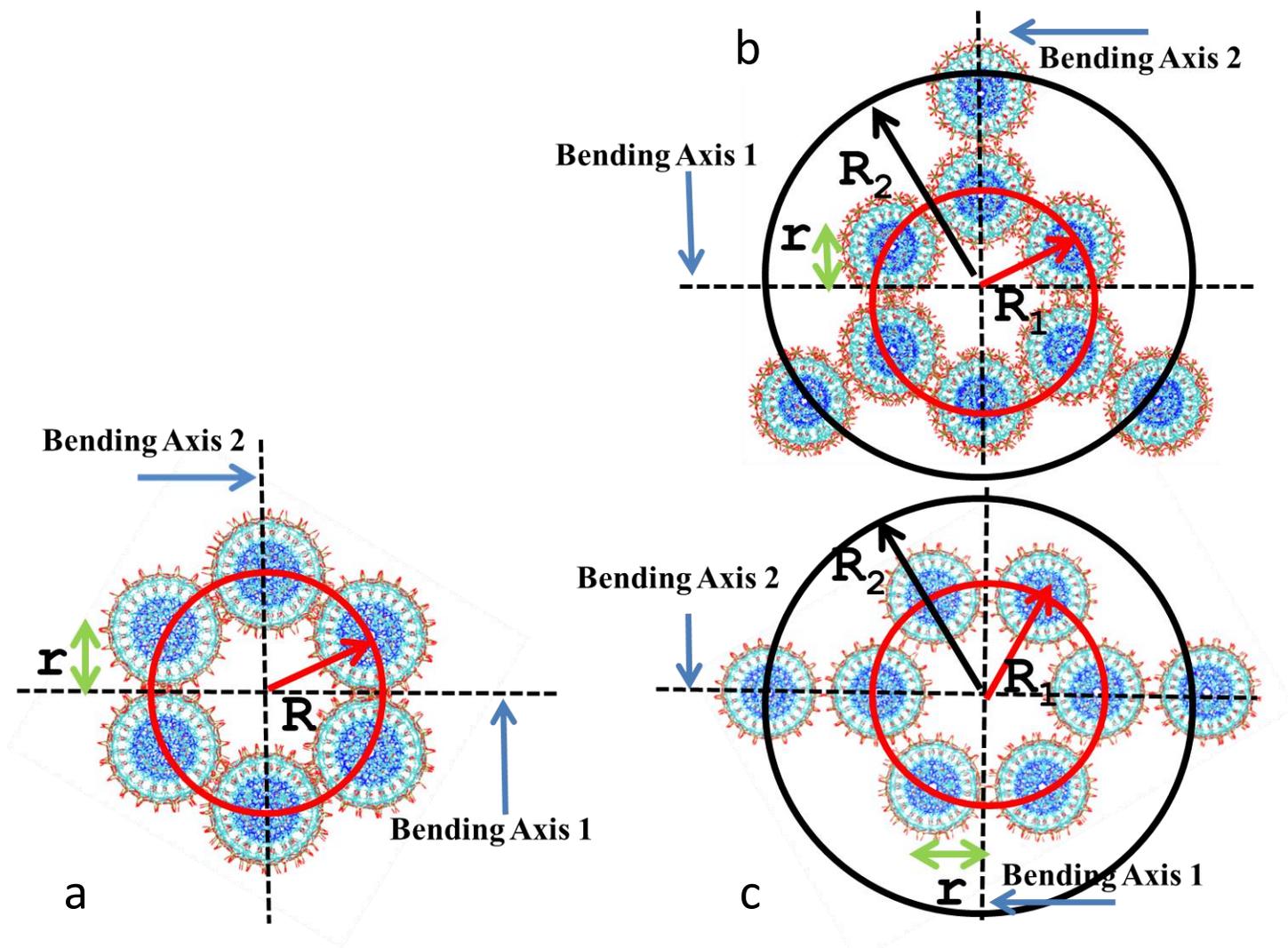

**Figure S7.**

**The cross-sectional view of the bending axis for (a) 6HB, (b) 6HB+3 and (c) 6HB+2. Radii of inner and outer circles, $R_1$ and $R_2$ have been taken to be 2.5 nm and 5.0 nm, respectively, by following the radius analysis of simulation data.**

The radius of the core of the DNT has been taken to be 2.5 nm and the radius including the outer helices has been taken to be 5.0 nm. These values have been taken from the central region of the equilibrated snapshots of DNT as shown in the radius analysis (Figure 4 and Figure S7).

If we consider DNA as an ideal cylindrical rod of radius 1 nm, the inner and outer radii of a DNT turns out to be 2 nm and 4 nm. This configuration will give different AMI. The values of the AMI with respect to the corresponding axis have been shown in the Table S2 (a). The AMI using ideal radii has been shown in the Table S2 (b) long with the corresponding values of AMI calculated by using the simulation radii.

**Table S2**

**The area moment of inertial of DNTs with respect to the two different bending axes. (a) AMI using the radii from the equlibrated structures in simulation (b) taking the ideal radii for helices and DNTs.**

(a)  $R_{inner}$=2.5 nm; $R_{outer}$ =5 nm

| Structure | $I_1$ ($\pi$ nm$^4$) | $I_2$ ($\pi$ nm$^4$) | $(I_1+I_2)/2$ ($\pi$ nm$^4$) |
|---|---|---|---|
| dSDNA | .25 | .25 | .25 |
| 6HB | 18 | 22.5 | 20.25 |
| 6HB+2 | 68.5 | 23 | 45.75 |
| 6HB+3 | 56.25 | 60.75 | 58.50 |

(b)  $R_{inner}$=2 nm; $R_{outer}$ = 4 nm

| Structure | $I_1$ ($\pi$ nm$^4$) | $I_2$ ($\pi$ nm$^4$) | $(I_1+I_2)/2$ ($\pi$ nm$^4$) |
|---|---|---|---|
| dSDNA | .25 | .25 | .25 |
| 6HB | 13.50 | 13.50 | 13.50 |
| 6HB+2 | 46 | 14 | 30 |
| 6HB+3 | 38.25 | 38.25 | 38.25 |

Where

$I_1$ = Area moment of inertia with respect to Bending axis 1

$I_2$ = Area moment of inertia with respect to Bending axis 2

$I$ = Average area moment of inertia

## 9. The Persistence Length of dSDNA from Bending Angles Distribution.

Small fluctuation in the bending angles provides a way to calculate the persistence length of a polymer like DNA. We plot the probability distribution of the bending angles of dSDNA in Figure S8 (a). The continuous line is a Gaussian fit to the distribution. The Gaussian nature of the probability distribution allows us to calculate the bending persistence length using the partition function resulting from a WLC model, also known as Kratky-Porod model.[8] The slope of the linear fit between 1-cos($\theta$) and ln P($\theta$) which is shown in Figure S8 (b), gives us the bending persistence length. We find a persistence length of 26.68 ($\pm$ 1.61) nm for the 12 mer dSDNA. The bending angle distribution of DNTs (shown in the Figure 9(a) in main manuscript) has a similar nature to the bending angle distribution of 12 mer dSDNA. This shows that the WLC model is suitable to model the mechanical strength of DNTs.

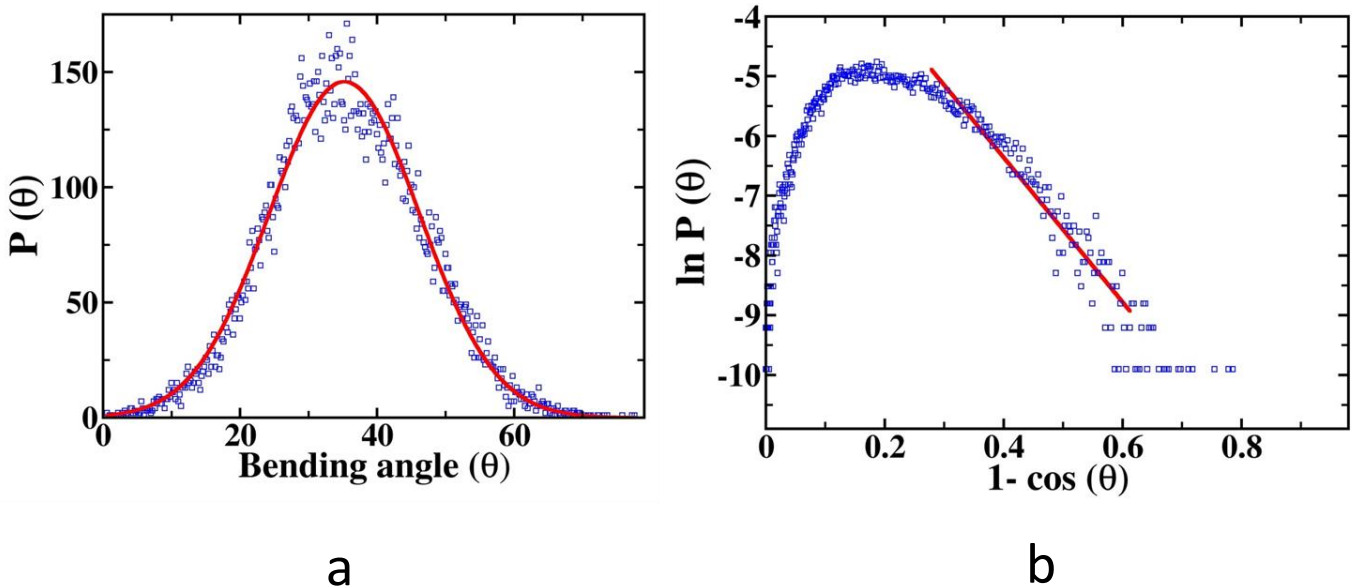

**Figure S8.**

**Probability distribution for the bend angle of 12-mer dSDNA in a 100 ns equilibrium MD simulation, (a) with respect to bending angle θ, (b) with respect to 1- cos (θ) on a semi log plot. The persistence length of dSDNA measured from this analysis is 26.68 (± 1.61) nm.**

## 10. The Persistence Length from Various Analyses.

In this study, the persistence lengths of DNTs have been characterized from various numerical schemes and approximations. Table S3 summarizes all the values of persistence length.

**Table S3. The Summary of Reported Persistence Length from various analyses.**

| Structure | SMD $P_l(\mu m)$ | SMD $P_l(\mu m)$ (Ideal radius) | Bending angle $P_l(\mu m)$ | Contour length $P_l(\mu m)$ | Experimental $P_l(\mu m)$ |
|---|---|---|---|---|---|
| dSDNA | .058 ± 0.003 | 0.057 (±.003) | 0.053 (±0.0004) | 0.057 (±.003) | 0.05 |
| 6HB | 5.3 (± 0.3) | 3.6 (±0.2) | 1.8 (±.1) | 4.9 (±0.2) | 1.0 (±0.1) |
| 6HB+2 | 12.1 (± 0.5) | 7.9 (± 0.3) | 2.6 (±0.1) | 13.5 (± 0.4) | 3.6 (±0.5) |
| 6HB+3 | 17.2 (± 0.5) | 11.2 (±0.3 ) | 2.9 (±0.1) | 21.9 (±0.7) | 5.0 (±0.5) |

## 11. The Snapshots of the DNTs after 200 ns simulations

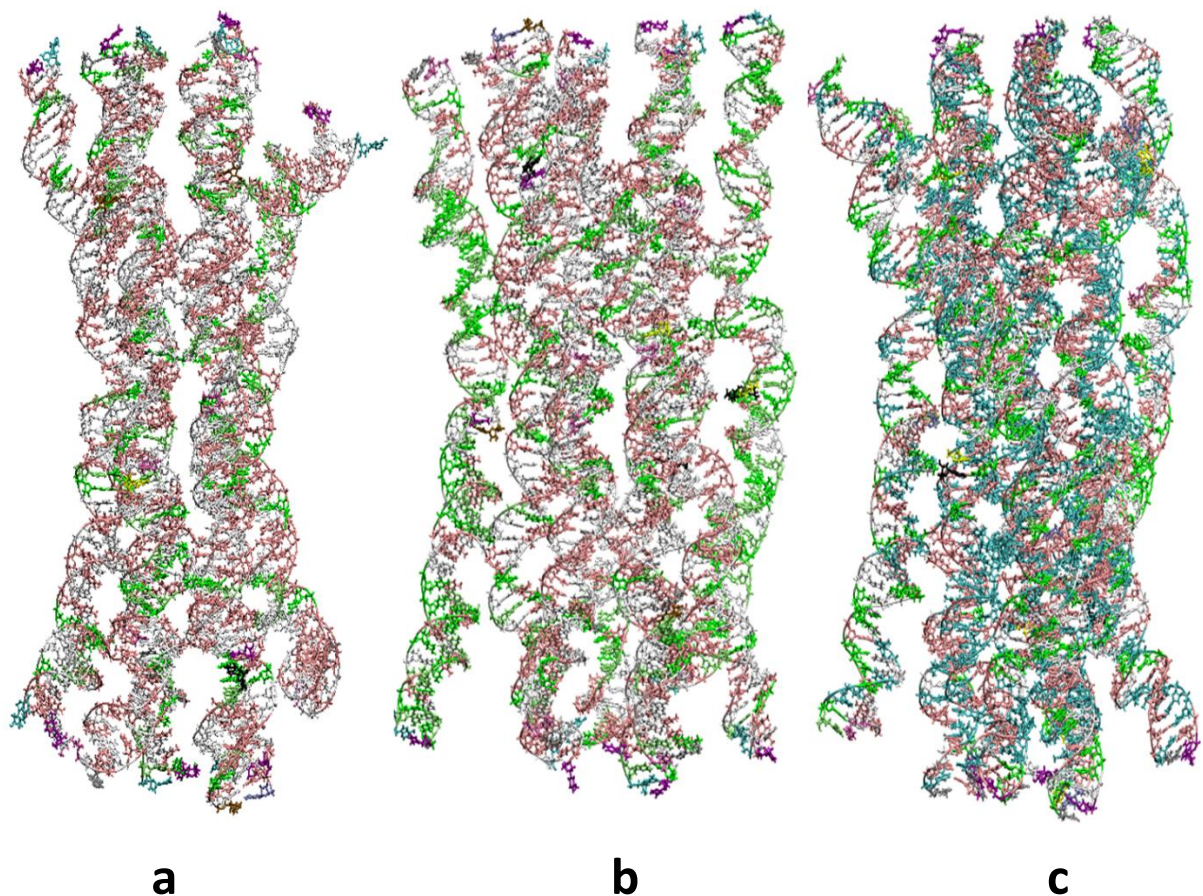

**Figure S9.**

Snapshots of (a) 6HB, (b) 6HB+2 and (c) 6HB+3 DNTs after 200 ns of MD simulation in "bond" representation.

## 12. The Analysis and Comparison of the Three Independent Simulation Trajectories.

In order to have better sampling of phase space, we have carried out three set of equilibrium MD simulation for each DNTs over 200 ns time scale. Figures S10 (a), (b) and (c) show the results for the simulations of 6HB, 6HB+2 and 6HB+3 respectively. The black, red and green corresponds to the simulation set 1, set 2 and set 3 respectively whereas the blue line show the average for the respective quantity. RMSD, RMSF, bending angle and contour length distribution are shown in the first second, third and fourth quadrant of each figure. Each snapshot was fit to the initially minimized structure before calculating RMSD and RMSF. The contour lengths and bending angles have been extracted from the last100 ns of three independent MD trajectories for each DNT. The average (blue curve) has been shown in the manuscript.

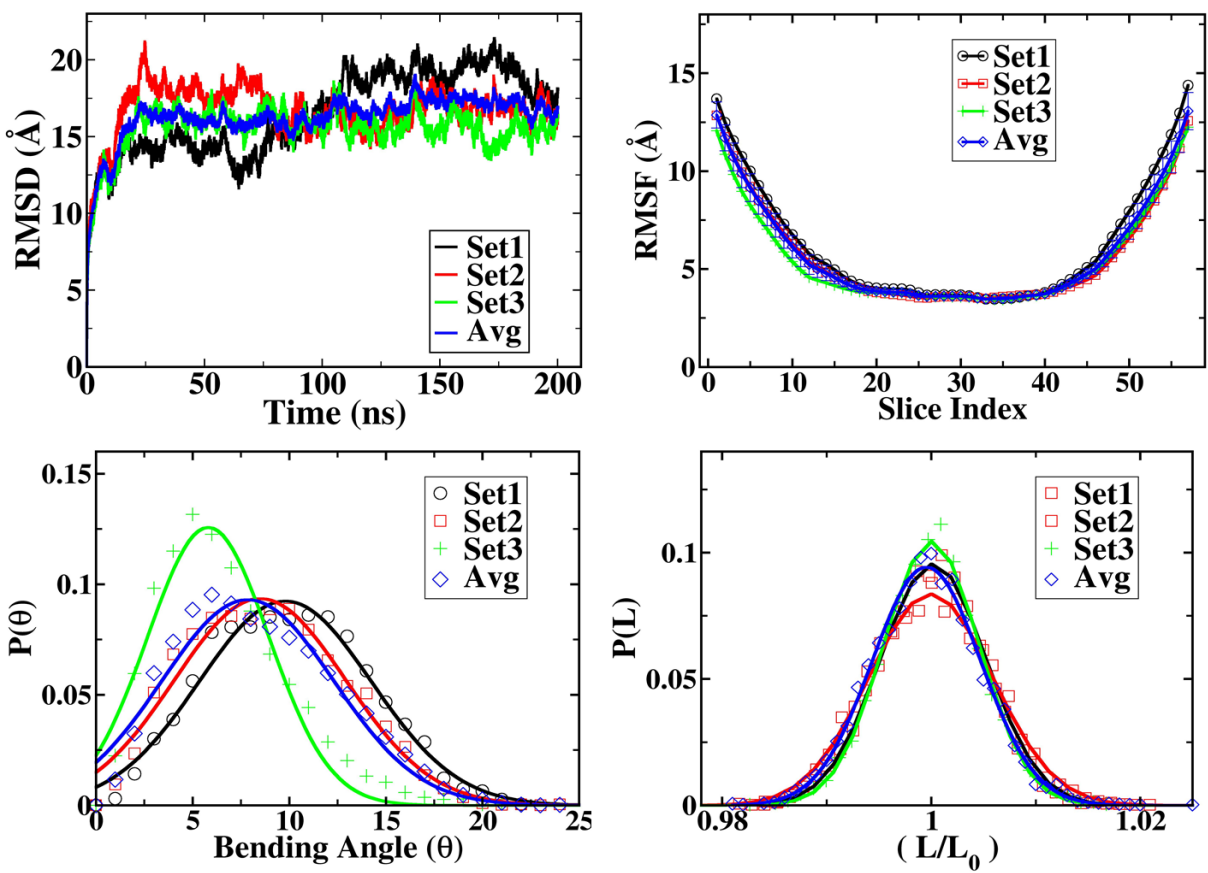

a

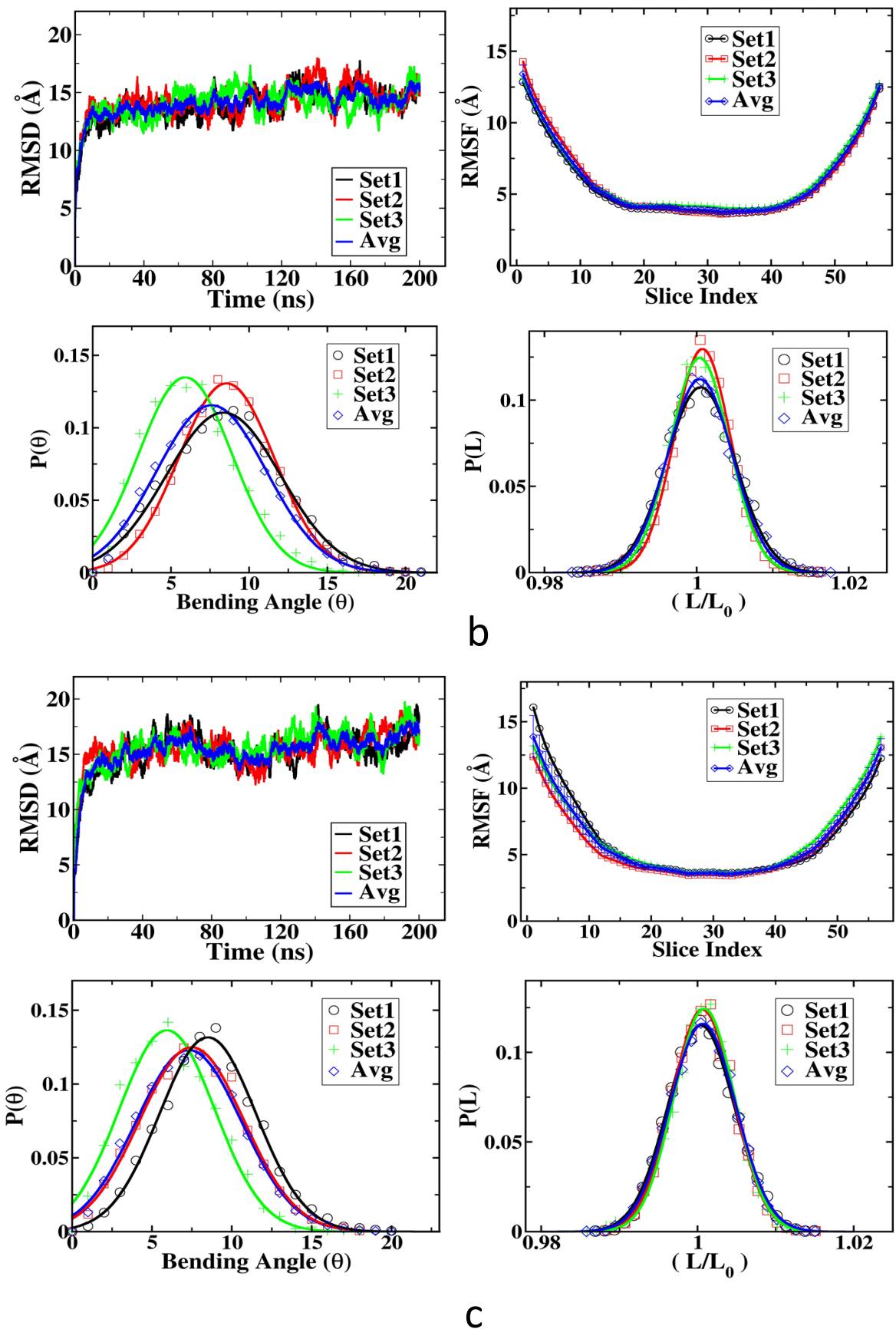

Figure S10. RMSD, RMSF, bending angle and contour length distributions of (a) 6HB, (b) 6HB+2 and (c) 6HB+3 DNA nanotubes.

## 13. The Structure with Original Sequence.

We have also carried the one set of simulations for 6HB, 6HB+2 and 6HB+3 DNA nanotubes with the exact sequence as described in the experiments. Figure S11 (a) shows the RMSD and RMSF for this set of simulation.

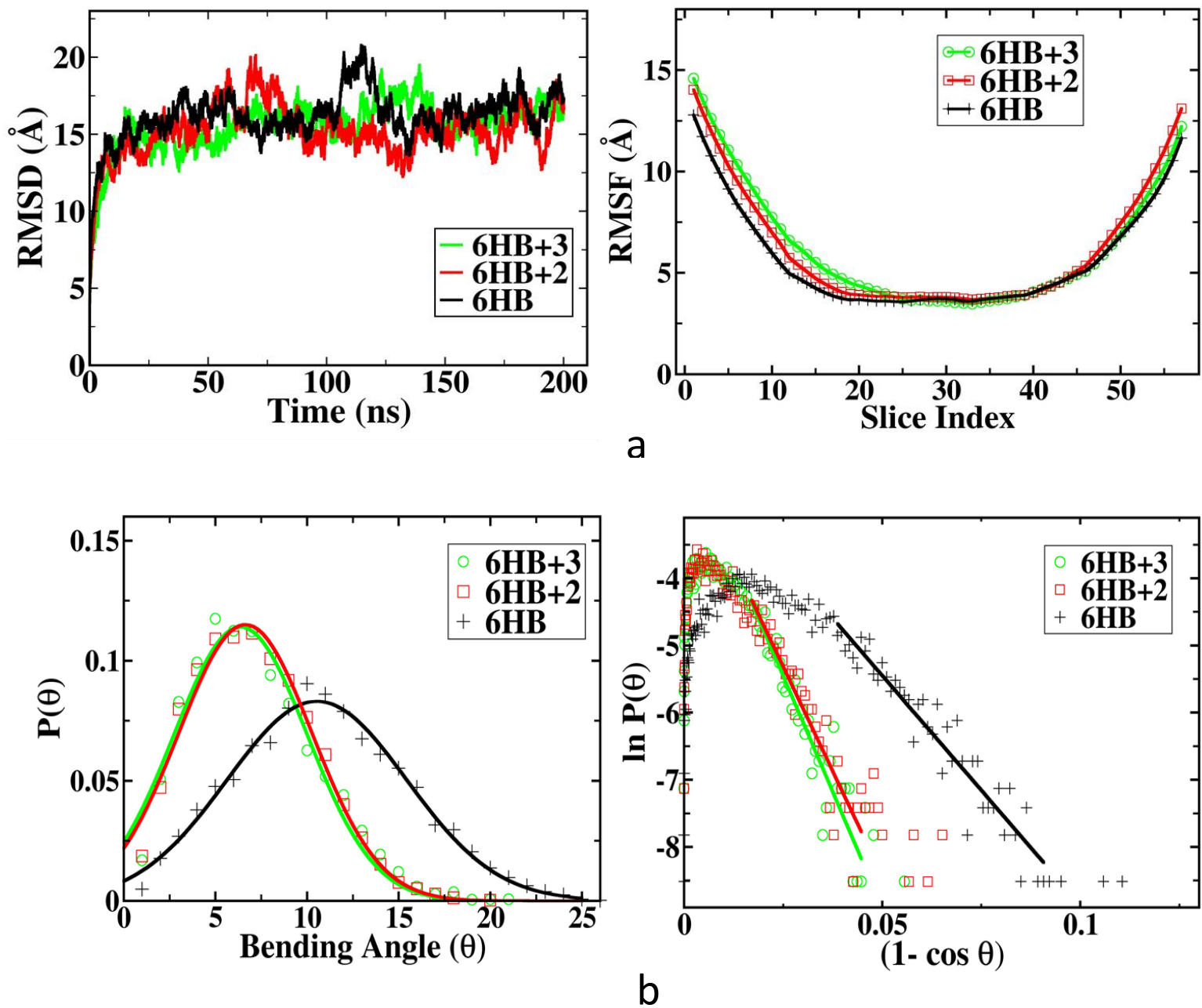

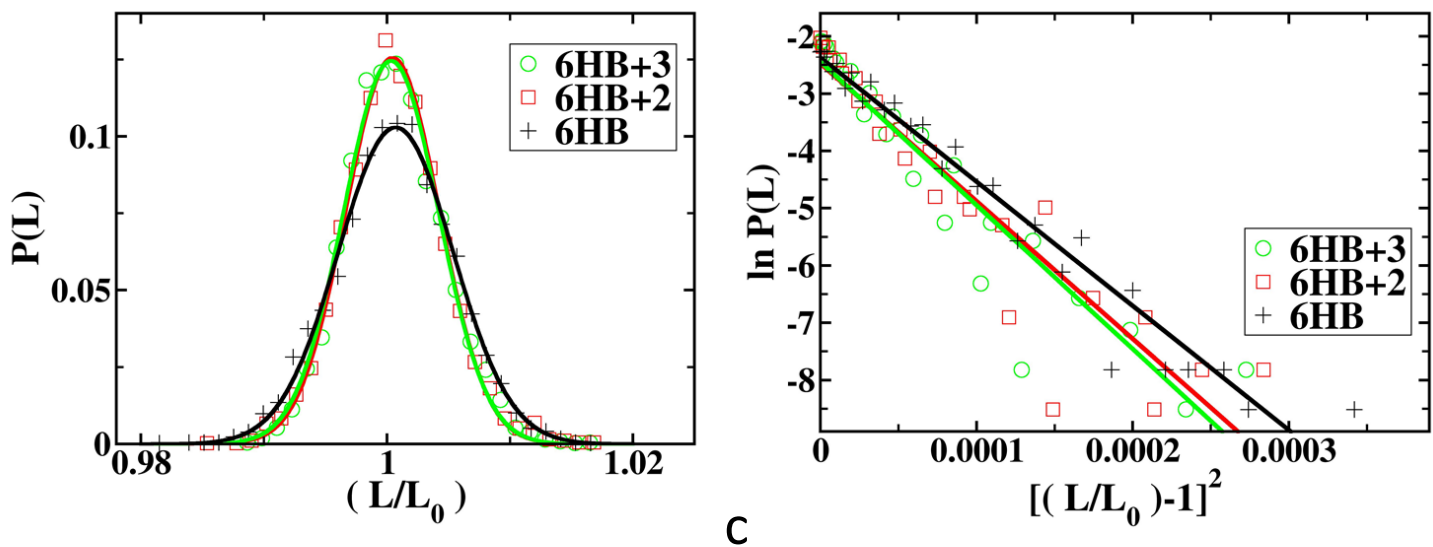

**Figure S11.**

(a) Comparison of the RMSD and RMSF values of 6HB, 6HB+2 and 6HB+3 DNTs with the experimental sequences over the course of 200 ns MD simulations. (b) Bending angle distribution of various DNTs from the last 100 ns of equilibrium MD simulation of the simulation trajectories. (c) The equilibrium contour length fluctuation of DNTs around their mean positions. The persistence length and stretch modulus has been obtained by the fitting of bending angles and contour lengths as shown in the figure.

The respective values of persistence length for 6HB, 6HB+2 and 6HB+3 are 1.41 (± 0.06), 2.56 (± 0.1) and 2.84 (± 0.1) μm. The stretch moduli extracted from the contour length distribution are 8621.5 (± 349.3), 9634.6 (±747.41) and 9999.6 (± 746.6) pN for 6HB, 6HB+2 and 6HB+3 respectively. The values are qualitatively similar to the corresponding values presented in the main manuscript with the sequence shown in the Figure S1. The difference is mostly within the computed error bars. It is imperative to state here that our simulations are not sensitive enough for gauging the effect of sequence on the mechanical strength of DNA nanotubes. Although this is an interesting question, we have not pursued it in this.

## 14. Stretch Modulus of B-DNA using SMD Simulation.

We have recorded the stress vs strain response of a 38mer B/DNA using an SMD protocol similar to the one used in this study. Figure S12 shows the force vs extension behavior of 38-mer BDNA from the constant velocity SMD simulation. This plot correctly reproduces various regions reported in various experimental studies.[9] We extract the stretch modulus using the Hooke's law from the linear region of the plot. The stretch modulus for this structure comes out to be 967.6 pN which is very similar to the reported values from experiments.[10]

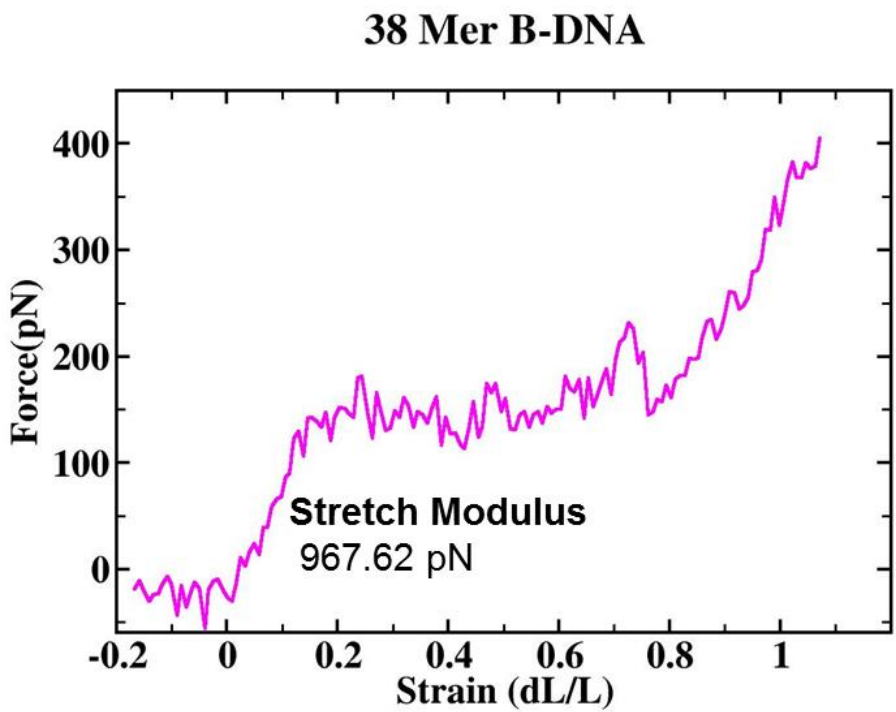

**Figure S12.**

**The force vs strain response of 38-mer B-DNA. We calculated the stretch modulus from the slope of the linear region of the plot.**

# REFERENCES

1. Macke, T. J.; Case, D. A. Modeling Unusual Nucleic Acid Structures. In *Molecular Modeling of Nucleic Acids*, Leontis, N. B.; SantaLucia, J., Eds. 1998; Vol. 682, pp 379-393.

2. D.A. Case, V. B., J.T. Berryman, R.M. Betz, Q. Cai, D.S. Cerutti, T.E. Cheatham, III, T.A. Darden, R.E. Duke, H. Gohlke, *et al*. Amber 14, University of California, San Francisco. **2014**.

3. Rich, A. DNA Comes in Many Forms. *Gene* **1993**, 135, 99-109.

4. Olson, W. K.; Bansal, M.; Burley, S. K.; Dickerson, R. E.; Gerstein, M.; Harvey, S. C.; Heinemann, U.; Lu, X. J.; Neidle, S.; Shakked, Z., et al. . A Standard Reference Frame for the Description of Nucleic Acid Base-Pair Geometry. *J. Mol. Biol.* **2001**, 313, 229-237.

5. Roe, D. R.; Cheatham, T. E., III. Ptraj and Cpptraj: Software for Processing and Analysis of Molecular Dynamics Trajectory Data. *J. Chem. Theory Comput.* **2013**, 9, 3084-3095.

6. Lu, X. J.; Olson, W. K. 3DNA: A Software Package for the Analysis, Rebuilding and Visualization of Three-Dimensional Nucleic Acid Structures. *Nucleic Acids Res.* **2003**, 31, 5108-5121.

7. Zgarbova, M.; Otyepka, M.; Sponer, J.; Lankas, F.; Jurecka, P. Base Pair Fraying in Molecular Dynamics Simulations of DNA and RNA. *J. Chem. Theory Comput.* **2014**, 10, 3177-3189.

8. Kratky, O.; Porod, G. Rontgenuntersuchung Geloster Fadenmolekule *Recl.: J. R. Neth. Chem. Soc.* **1949**, 68, 1106-1122.

9. Marko, J. F.; Siggia, E. D. Stretching DNA. *Macromolecules* **1995**, 28, 8759-8770.

10. Smith, S. B.; Cui, Y. J.; Bustamante, C. Overstretching B-DNA: The Elastic Response of Individual Double-Stranded and Single-Stranded DNA Molecules. *Science* **1996**, 271, 795-799.